\documentclass[twocolumn,twocolappendix]{openjournal}
\usepackage{amsmath}
\usepackage{amssymb}
\usepackage{bm}
\usepackage{graphicx}

\usepackage[breaklinks=true]{hyperref}
\hypersetup{
colorlinks=true, 
allcolors=blue, 
bookmarksopen=true,
bookmarksopenlevel=1,
bookmarksnumbered=true,
pdfstartview={FitH},
pdftitle={Polarization Properties of Thomson-Scattering Triaxial Ellipsoids},
pdfauthor={Naoki Hachisu, Tatsuya Matsumoto and Tomonori Totani}
}
\usepackage{orcidlink}

\usepackage{setspace}
\shortauthors{Hachisu et al.}

\begin{document}

\title{Polarization Properties of Thomson-Scattering Triaxial Ellipsoids}

\author{Naoki Hachisu$^1$ \orcidlink{0009-0007-9960-9818}}
\author{Tatsuya Matsumoto$^1$ \orcidlink{0000-0002-9350-6793}}
\author{Tomonori Totani$^1$ \orcidlink{0000-0002-6156-6783}} \vspace{1pt}
\affiliation{$^1$ Department of Astronomy, Graduate School of Science, the University of Tokyo, Bunkyo-ku, Tokyo 113-0033, Japan}

\begin{abstract}
Polarimetry is a powerful tool for constraining the geometry of distant astrophysical objects. Continuum polarization induced by Thomson scattering can probe the photospheric geometry of transients and provide insight into their explosion mechanisms. We systematically study continuum polarization from spheroids over a broader range of deformations and viewing angles than considered previously, and generalize the calculation to triaxial ellipsoids. We find that strongly deformed spheroids exhibit a non-monotonic dependence of polarization on viewing angle, a behavior overlooked in previous studies. Triaxial ellipsoids show more diverse polarization signatures, but their qualitative behavior can be understood from the area of surface elements viewed nearly edge-on, where the local polarization is strongest. Our results provide a simple physical understanding of the dependence of polarization on photospheric geometry and help constrain the geometry of astrophysical transients through polarimetric observations. The computational results of this study will be made publicly available for use in future research.
\end{abstract}

\section{Introduction}\label{sec:intro}
Polarization provides a powerful probe of the geometry of distant unresolved astrophysical transients \citep[e.g.,][]{Wang&Wheeler2008}. In optical transients powered by thermal emission, such as supernovae (SNe), photons become linearly polarized through Thomson scattering as they diffuse outward through the ejecta. The observed continuum polarization is obtained by integrating the local polarization over the entire photosphere \citep{Chandrasekhar1960,Shapiro&Sutherland1982,Hoflich1991}. While the net polarization vanishes for a spherical photosphere, any departure from spherical symmetry produces a finite polarization signal, thereby providing direct constraints on the geometry of the emitting region.

Continuum polarization has been most extensively studied in SNe \citep{Wang&Wheeler2008,Patat2017}. During the photospheric phase, most SNe exhibit modest continuum polarization, typically below $\lesssim1\,\%$, indicating only moderate departures from spherical symmetry \citep{Wang+1996b,Leonard&Filippenko2001,Leonard+2001,Wang+2003b,Leonard+2005,Cikota+2019,Nagao+2024}. Among core-collapse SNe, the polarization frequently increases as the photosphere recedes into deeper ejecta layers toward the onset of the nebular phase \citep{Leonard+2006,Chornock+2010b,Kumar+2016,Nagao+2024}. Likewise, stripped-envelope SNe generally exhibit relatively larger continuum polarization than hydrogen-rich events \citep{Wang+2001,Maund+2007,Tanaka+2008c,Tanaka+2009b}. These observations suggest that the inner ejecta are more aspherical, possibly reflecting an explosion mechanism \citep[e.g.,][]{Wang+2001}. In addition, SNe interacting with dense circumstellar material often exhibit exceptionally large continuum polarization of $\gtrsim1\,\%$ \citep{Leonard+2000,Wang+2001,Mauerhan+2024,Bilinski+2024}, indicating a highly aspherical circumstellar environment.

More recently, polarimetric observations have expanded to very early phases of SN evolution \citep{Vasylyev+2023,Singh+2024,Shrestha+2025,Vasylyev+2026}, as well as to a broader class of optical transients such as superluminous SNe \citep{Inserra+2016b,Pursiainen+2023}, fast blue optical transients \citep{Maund+2023}, and tidal disruption events \citep[TDEs,][]{Wichern+2026}. In particular, thermal TDEs typically exhibit polarization of $\simeq1-3\,\%$ \citep{Patra+2022,Leloudas+2022,Koljonen+2024c,JordanaMitjans+2025,Koljonen+2025b,Uno+2025}, while a few events have shown even higher polarization \citep{Lee+2020,Liodakis+2023,Floris+2025b}. These observations highlight the growing importance of polarization as a diagnostic of the geometry and emission mechanisms of optical transients.

Despite the growing number of polarimetric observations, most theoretical studies have focused on axisymmetric geometries \citep{Shapiro&Sutherland1982,Hoflich1991,Dessart&Hillier2011b,Dessart+2024b,Nagao+2026}, and the polarization signatures of more general photospheric configurations remain largely unexplored. In this work, adopting the analytical framework of \cite{Shapiro&Sutherland1982}, we investigate the polarization from general triaxial ellipsoids as a first step toward understanding polarization signatures from non-axisymmetric photospheres. This paper is organized as follows. In Section~\ref{sec:method}, we describe the method used to calculate polarization from aspherical photospheres. In Section~\ref{sec:result}, we present results for spheroidal geometries over a broader parameter space than considered in previous studies, and then extend the analysis to general triaxial ellipsoids. Our findings are summarized in Section~\ref{sec:summary}.

\vspace{1.0\baselineskip}
\section{Method} \label{sec:method}

\begin{figure*}[t!]
\centering
\includegraphics[width=\textwidth]{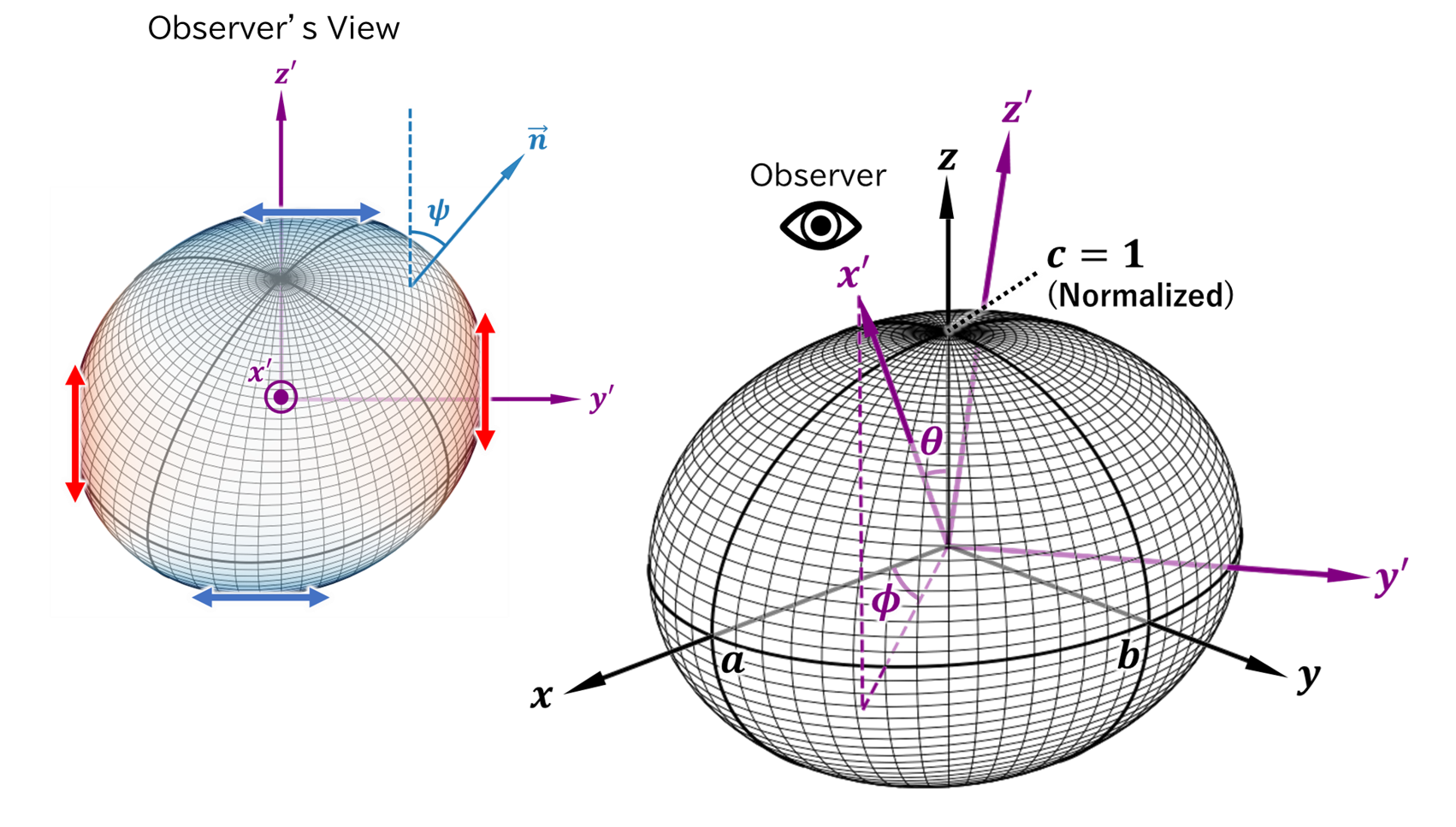}
\caption{Schematic illustration of the ellipsoidal photosphere and the viewing configuration. The ellipsoid has its principal axes aligned with the $x$, $y$, and $z$-axes. The primed coordinate system is defined such that the $x^\prime$-axis points toward the observer, and the $y^\prime z^\prime$-plane corresponds to the sky plane. Surface elements whose projected normal vectors are aligned with the $y^\prime$ ($z^\prime$) direction produce polarization aligned with the $z^\prime$ ($y^\prime$) direction, illustrated by the red (blue) local polarization vectors.}
\label{fig:picture}
\end{figure*}

We compute the polarization signal emerging from an ellipsoidal photosphere following the method of \cite{Shapiro&Sutherland1982}. The polarization state of radiation is fully described by the four Stokes parameters, $I$, $Q$, $U$, and $V$ \citep[e.g.,][]{Rybicki&Lightman1979}. For an unresolved source, the observed (hereafter ``net'') polarization is obtained by integrating the Stokes parameters over the entire visible surface.

The local contribution from each surface element is calculated using the classical solution of \cite{Chandrasekhar1960} for a semi-infinite, plane-parallel atmosphere dominated by Thomson scattering. Owing to the plane-parallel geometry, the emergent polarization depends only on the viewing angle measured from the surface normal. The polarization direction is perpendicular to the plane defined by the surface normal and the line of sight (the so-called meridian plane), and the polarization degree increases with viewing angle, reaching a maximum of $\simeq11.7\,\%$ at a viewing angle of $90^\circ$. Choosing the reference axis for the Stokes parameters to be parallel to the meridian plane, the Stokes parameters satisfy $Q\leq0$ and $U=0$, with the latter following from reflection symmetry. We also have $V=0$ because Thomson scattering does not generate circular polarization from initially unpolarized radiation. We adopt the flux-normalized Chandrasekhar solution for $I(\mu)/F$ and $Q(\mu)/F$ in tabulated form (see, e.g., Table~1 of \citealt{Shapiro&Sutherland1982}). Here $\mu$ is the cosine of the viewing angle measured from the surface normal, and $F$ is defined as
\begin{align}
F = 2\int^{1}_{0} I(\mu)\,\mu\,d\mu \ ,
\end{align}
so that $\pi F$ corresponds to the total emergent radiative flux per unit surface area. As discussed below, we assume that $F$ is uniform over the emitting surface.

In this work, we consider general triaxial ellipsoids as the shape of the emitting surface, rather than the spheroidal geometries originally considered by \cite{Shapiro&Sutherland1982}. In a Cartesian coordinate system, the surface is described by
\begin{equation}
\frac{x^2}{a^2} + \frac{y^2}{b^2} + \frac{z^2}{c^2} = 1\ ,
    \label{eq:ellipsoid}
\end{equation}
where $a$, $b$, and $c$ are the semi-axis lengths, defined up to an overall scaling. Without loss of generality, we set $c = 1$ hereafter. The observer views the ellipsoid along a direction specified by a unit vector $\hat{\bm{x}}^\prime$, whose polar and azimuthal angles are $\theta$ and $\phi$, respectively. Due to the symmetry of the ellipsoidal shape, it is sufficient to restrict the angles to $0 \leq \theta, \phi \leq \pi/2$. Figure~\ref{fig:picture} shows a schematic illustration of the ellipsoidal geometry and the viewing configuration.

The net polarization is obtained by integrating the locally defined Stokes parameters over the visible surface after transforming them into a common observer frame. For this purpose, we introduce a rotated Cartesian coordinate system $(x^\prime, y^\prime, z^\prime)$, with the $x^\prime$-axis aligned with the line-of-sight direction $\hat{\bm{x}}^\prime$. The $y^\prime$- and $z^\prime$-axes span the plane of the sky, and the $z^\prime$-axis is chosen to lie in the $x^\prime z$-plane (see Fig.~\ref{fig:picture}).

At each surface element, the local polarization is described by two orthogonally polarized intensity components, $I_l$ and $I_r$, whose electric vectors are parallel and perpendicular to the local meridian plane, respectively. These intensities are related to the Stokes parameters by
\begin{align}
I &= I_l + I_r\ ,\\
Q &= I_l - I_r\ .
\end{align}
These local polarization components are transformed into the observer frame as
\begin{align}
I^\prime_l &= I_l\cos^2\psi + I_r\sin^2\psi\ ,\\
I^\prime_r &= I_l\sin^2\psi + I_r\cos^2\psi\ ,\\
U^\prime &= (I_l - I_r)\sin 2\psi\ ,
\end{align}
where the primed quantities are defined with respect to the reference plane spanned by $\hat{\bm{x}}^\prime$ and $\hat{\bm{z}}^\prime$, such that $I^\prime_l$ and $I^\prime_r$ denote the intensities polarized parallel and perpendicular to this plane, respectively. The rotation angle $\psi$ is defined as the angle between the $\hat{\bm{z}}^\prime$-axis and the projection of $\hat{\bm{n}}$ onto the sky ($y^\prime z^\prime$) plane:\footnote{We note that Eq.~(27) of \cite{Shapiro&Sutherland1982} contains an error, because the projected surface normal onto the $yz$-plane, $\hat{\bm{x}}^\prime\times(\hat{\bm{n}}\times\hat{\bm{x}}^\prime)$, is not properly normalized.}
\begin{align}
\psi=\arccos\left[\frac{\hat{\bm{z}}^\prime\cdot\left\{\hat{\bm{x}}^\prime\times(\hat{\bm{n}}\times\hat{\bm{x}}^\prime)\right\}}{\sqrt{1-(\hat{\bm{n}}\cdot\hat{\bm{x}}^\prime)^2}}\right]\ .
\end{align}
The Stokes parameters in the observer frame are then given by
\begin{align}
I^\prime &= I_l^\prime + I_r^\prime = I_l + I_r = I\ ,
    \label{eq:Iprime}\\
Q^\prime &= I_l^\prime - I_r^\prime = (I_l-I_r)\cos2\psi = Q\cos2\psi\ ,
    \label{eq:Qprime}\\
U^\prime &= Q\sin2\psi\ ,
    \label{eq:Uprime}
\end{align}
where $I$ and $Q$ are given by the Chandrasekhar solution and depend solely on $\mu=\hat{\bm{n}}\cdot\hat{\bm{x}}^\prime$ up to the normalization factor $F$.  The circular polarization remains zero under this transformation, $V^\prime=V=0$.

The observed net Stokes parameters are obtained by averaging the local contributions (Eqs.~\ref{eq:Iprime}-\ref{eq:Uprime}) over the projected visible surface. Using the flux-normalized Chandrasekhar solution, they can be written as:
\begin{align}
I^\prime_{\rm net}&=\frac{1}{A_p}\int dA_p\,\left[ \frac{I(\mu)}{F} \right] F\ ,
    \label{eq:Inet}\\
Q^\prime_{\rm net}&=\frac{1}{A_p}\int dA_p\,\cos 2\psi\,\left[ \frac{Q(\mu)}{F} \right] F\ ,
    \label{eq:Qnet}\\
U^\prime_{\rm net}&=\frac{1}{A_p}\int dA_p\,\sin 2\psi\,\left[ \frac{Q(\mu)}{F} \right] F\ ,
    \label{eq:Unet}
\end{align}
where $dA_p$ is the projected area element of a surface element $dA$ onto the sky plane:\footnote{For the ellipsoidal geometry, the visible area is given by $\mu\geq0$.}
\begin{align}
dA_p=\mu dA\ .
\end{align}
The total projected area is correspondingly given by
\begin{align}
A_p=\int dA_p\ .
\end{align}

In general, the surface flux $F$ may vary across a non-spherical photosphere. In the present work, our primary aim is to investigate the dependence of polarization on the global geometry, while modeling the detailed flux distribution over the emitting surface is beyond the scope of this study. As a first step, we therefore assume a uniform surface flux, $F=const$.

Finally, the net polarization degree and polarization angle are obtained from the integrated Stokes parameters as
\begin{align}
\Pi&=\frac{\sqrt{(Q_{\rm net}^\prime)^2+(U_{\rm net}^\prime)^2}}{I_{\rm net}^\prime}\ ,
    \label{eq:Pi}\\
\chi&=\frac{1}{2}\arctan\left(\frac{U_{\rm net}^\prime}{Q_{\rm net}^\prime}\right)\ .
    \label{eq:chi}
\end{align}
Here the polarization angle specifies the orientation of the polarization vector on the sky plane and is defined modulo $\pi$, with $\chi=0$ and $\chi=\pm\pi/2$ corresponding to polarization aligned with the $z^\prime$- and $y^\prime$-axes, respectively. We adopt the range $-\pi/2<\chi\leq\pi/2$. Note that the assumed constant flux $F$ and the projected surface area $A_p$ cancel out in the calculation of the polarization degree and angle.

\vspace{1.0\baselineskip}
\section{Results} \label{sec:result}
We calculate the polarization from triaxial ellipsoids. The shape of an ellipsoid is characterized by the two axis ratios, $a$ and $b$, which are varied over the range $0.1\leq a,b\leq10$. The viewing direction is specified by the polar and azimuthal angles, $\theta$ and $\phi$, respectively. Therefore, the polarization properties are fully determined by these four parameters.

\subsection{Spheroids}
\label{sec:spheroids}
We begin by examining the polarization properties of spheroids ($a=b$) before discussing the general triaxial case. The spheroidal model was originally studied by \cite{Shapiro&Sutherland1982}; however, they evaluated the polarization only for equatorial viewing directions ($\theta=\pi/2$). Here, we extend their analysis and calculate the polarization for a general angle and investigate its dependence on both the viewing angle and the geometry of the emitting surface. Owing to the axisymmetry of spheroids, the Stokes parameter $U_{\rm net}^\prime$ vanishes identically, and Eq.~\eqref{eq:Pi} reduces to 
\begin{align}
\Pi = \frac{|Q_{\rm{net}}^\prime|}{I_{\rm net}^\prime}\ .
    \label{eq:Pi2}
\end{align}
We also confirm that our results for the special viewing angle $\theta=\pi/2$ are consistent with those reported by \cite{Shapiro&Sutherland1982}.

\begin{figure}[t!]
\centering
\includegraphics[width=0.45\textwidth, trim=2pt 2pt 2pt 2pt, clip]{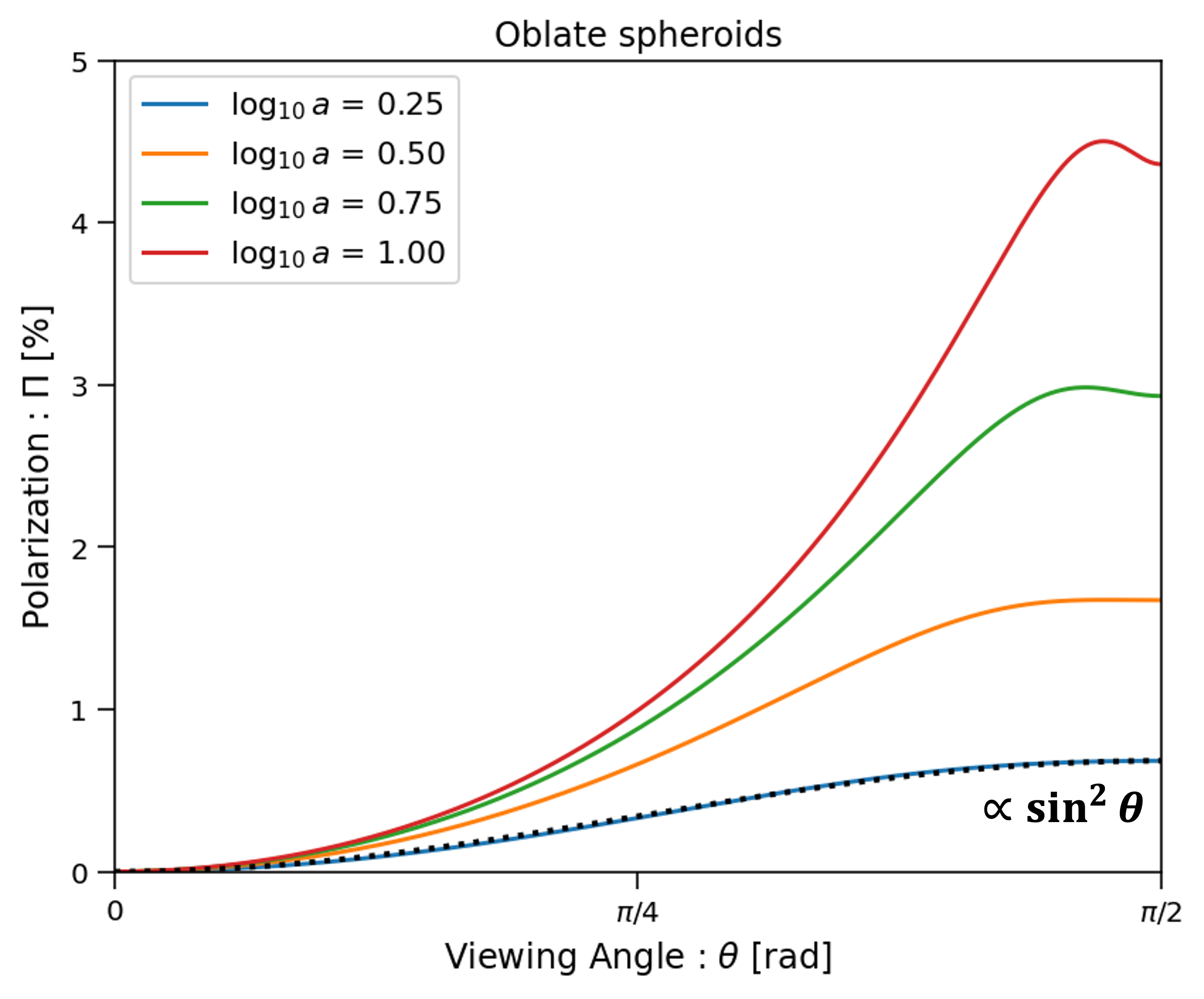}
\caption{The degree of polarization of oblate spheroids as a function of viewing angle, $\theta$, for different values of the semi-axis lengths (ratios), $a$. The black dotted curve shows the analytical scaling $\Pi \propto \sin^2\theta$, which captures the polarization behavior for moderate values of $a$.}
\label{fig:oblate}
\end{figure}

Figure~\ref{fig:oblate} shows the polarization degree as a function of the viewing angle for oblate spheroids, i.e., geometries compressed along the symmetry axis ($a=b>1$). The polarization degree generally increases with both viewing angle and oblateness. This viewing-angle dependence can be understood geometrically \citep[e.g.,][]{Kasen+2003}. For a pole-on view ($\theta=0$), the polarization vanishes due to symmetry. In an oblate spheroid, the broad upper surface occupies a large fraction of the projected image. As the viewing angle increases, this surface is viewed progressively closer to edge-on, where the local polarization is strongest. At larger viewing angles, part of the lower surface also becomes visible and contributes in the same manner. The projected normal vectors of these upper and lower surfaces are predominantly aligned with the $z^\prime$-axis, so that their polarization is preferentially oriented along the $y^\prime$-axis. As a result, the net polarization generally increases with $\theta$, while the polarization angle remains fixed at $\chi=\pi/2$.

 For mildly oblate configurations, the angular dependence is approximately described by $\Pi\propto\sin^2\theta$ \citep{Brown&McLean1977,Hoflich1991}. Interestingly, for sufficiently large oblateness ($a\gtrsim2.9$ or $\log a\gtrsim0.46$), the polarization no longer increases monotonically with viewing angle, but instead reaches a maximum at an angle slightly smaller than $\theta=\pi/2$. This behavior was not identified by \cite{Shapiro&Sutherland1982}, who implicitly assumed that the maximum polarization is attained at $\theta=\pi/2$. The slight reduction toward the exact equatorial view is caused by a small decrease in the surface area contributing to polarization along the $y^\prime$-axis.

\begin{figure}[t!]
\centering
\includegraphics[width=0.45\textwidth, trim=2pt 2pt 2pt 2pt, clip]{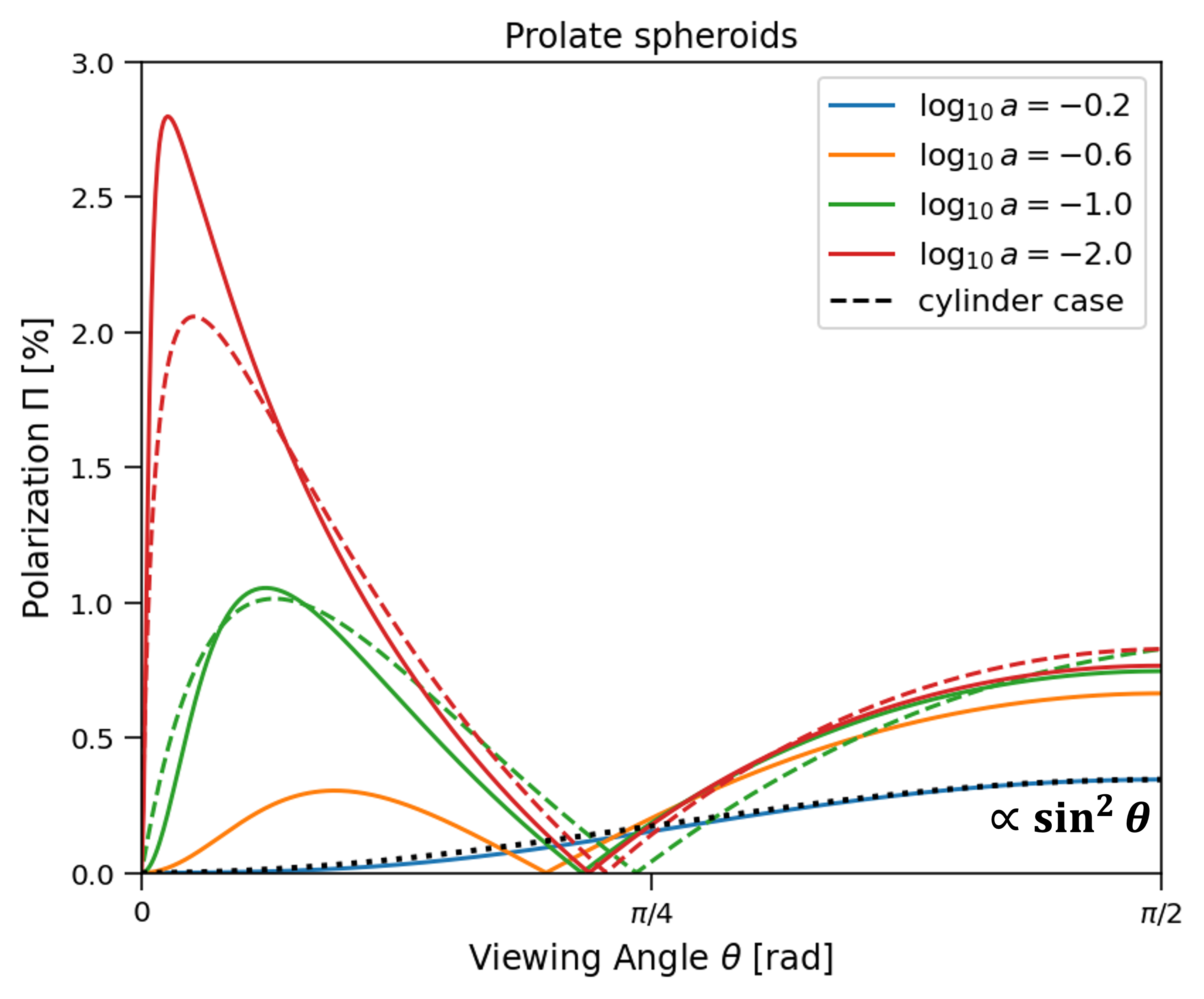}
\caption{Same as Fig.~\ref{fig:oblate}, but for prolate spheroids. In this regime, we define $\theta_{\rm null}$ as the viewing angle at which the degree of polarization vanishes between $\theta=0$ and $\pi/2$. The dashed curves for $a\leq0.1$ show the polarization for a cylindrical geometry derived in Appendix~\ref{appendix}.}
\label{fig:prolate}
\end{figure}

\begin{figure*}
\centering
\includegraphics[width=\columnwidth, trim=2pt 2pt 2pt 2pt, clip]{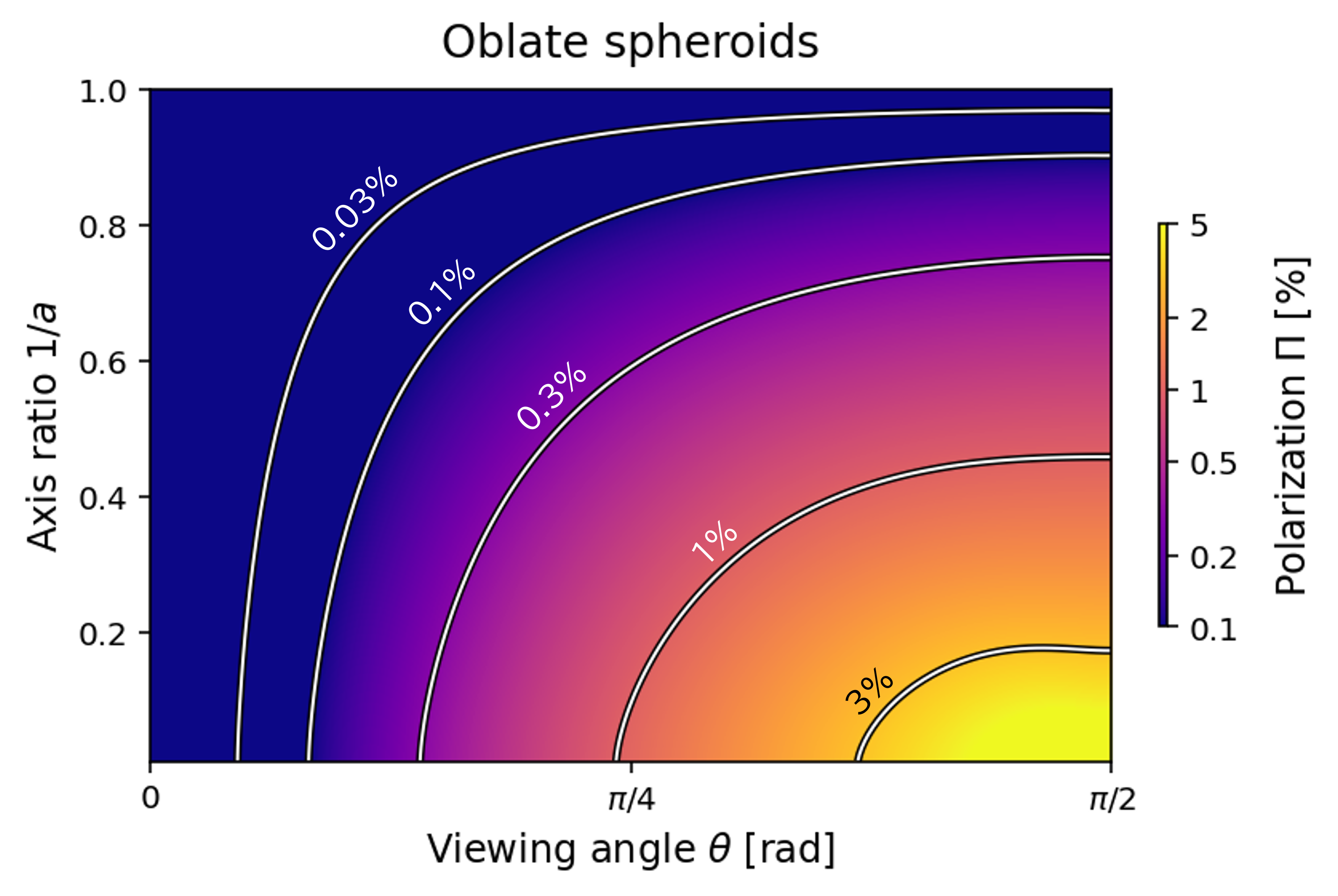}
\includegraphics[width=\columnwidth, trim=2pt 2pt 2pt 2pt, clip]{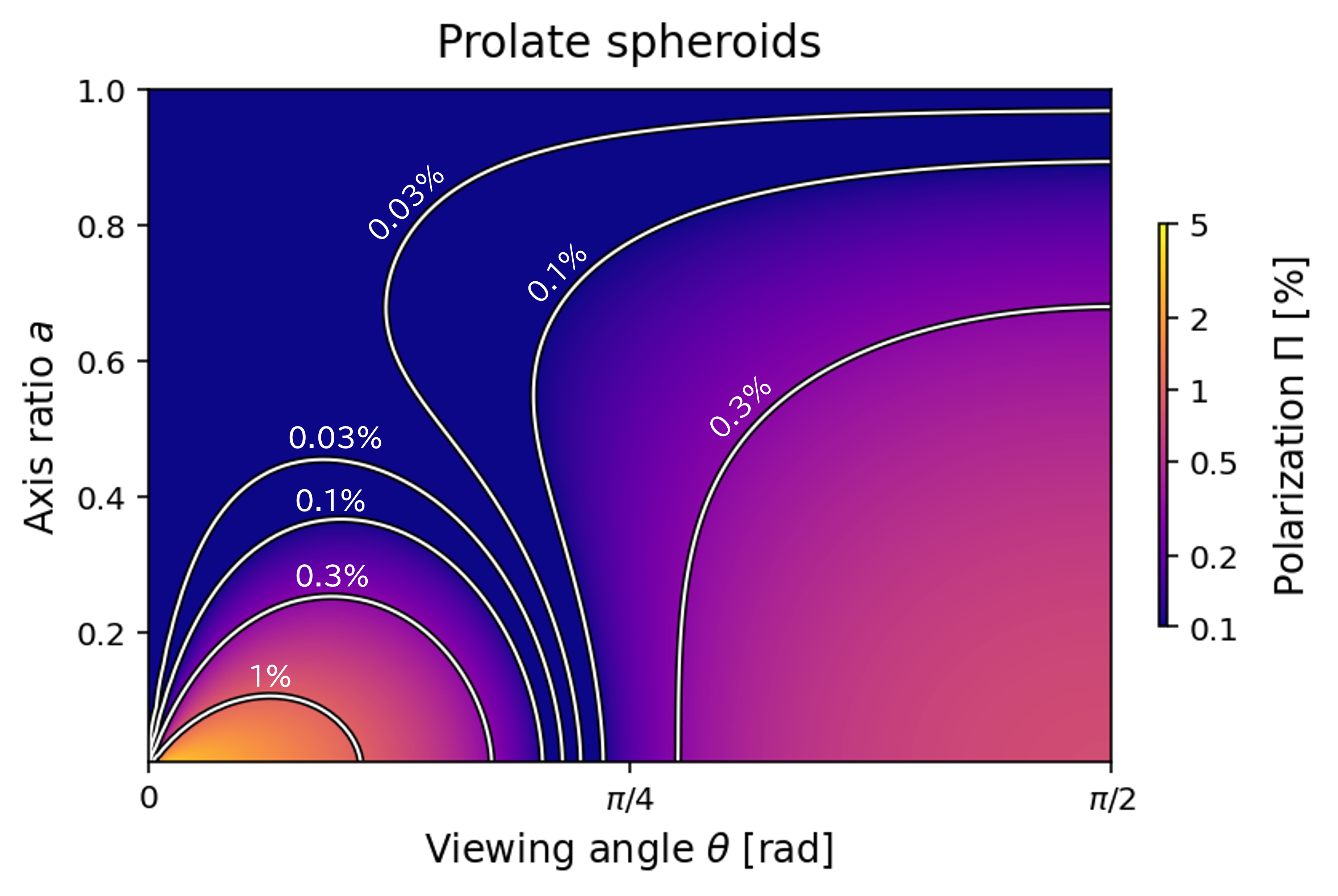}
\caption{The degree of polarization as a function of viewing angle and axis ratio. The left and right panels show the distributions for oblate and prolate spheroids, respectively. Polarization degrees outside the range $0.1\text{–}5\,\%$ are clipped to the corresponding limits of the color scale. For oblate spheroids, we define the axis ratio as $1/a$ to facilitate comparison with previous works (e.g., figure 3 of \citealt{Matsumoto2018}).
}
\label{fig:polarization map spheroid}
\end{figure*}

Figure~\ref{fig:prolate} shows the polarization degree for prolate spheroids, i.e., geometries elongated along the symmetry axis ($a=b<1$). The polarization degree exhibits more complicated dependence on the viewing angle than the oblate case. For a mild departure from sphericity ($0.57\lesssim a<1$ or $-0.24\lesssim \log a<0$), the behavior is similar to that of oblate spheroids with $\Pi\propto \sin^2\theta$. However, for highly elongated geometries, the polarization exhibits a double-peaked structure and vanishes at an intermediate viewing angle, $\theta_{\rm null}$. For these highly elongated prolate spheroids, the behavior at small viewing angles is initially similar to that of oblate spheroids. The upper surface dominates the polarization, and the polarization degree increases from zero as the viewing angle increases. However, the elongated side surface soon begins to contribute significantly and produces polarization with an orthogonal polarization angle. As its contribution grows, it increasingly cancels the polarization from the upper surface, causing the net polarization to decrease and eventually vanish at $\theta_{\rm null}$. Beyond this angle, the side-surface contribution dominates, and the polarization angle flips by $90^\circ$, from $\chi=\pi/2$ to $\chi=0$. This competition between the upper and side surfaces produces the characteristic double-peaked polarization structure. This non-monotonic behavior was not identified by \cite{Shapiro&Sutherland1982}.

We find that the behavior of a highly-prolate spheroid's polarization can be qualitatively understood by approximating the spheroid as a cylinder. In Appendix~\ref{appendix}, we analytically calculate the polarization from a cylindrical geometry, and compare the result with the numerical calculation for prolate spheroids in Fig.~\ref{fig:prolate}. In the extreme limit $a\to0$, the numerical result approaches the analytical cylindrical solution. In this limit, the null angle converges $\theta_{\rm null}\simeq0.226\pi\simeq41^\circ$.

In several previous studies aiming to constrain the photospheric geometry and viewing angle from continuum polarization, the analytical scaling $\Pi\propto \sin^2\theta$ has been adopted for both oblate and prolate spheroids \citep[e.g.,][]{Matsumoto2018,Uno+2023b}, motivated by earlier studies \citep{Brown&McLean1977,Hoflich1991}. However, as explicitly demonstrated above, this simple scaling breaks down for sufficiently large deviations from spherical symmetry. In particular, for prolate geometries, significant deviations already appear for relatively modest elongations. Figure~\ref{fig:polarization map spheroid} presents contour maps of the polarization degree as functions of viewing angle and axis ratio. While the results for the oblate case are qualitatively consistent with the distribution assumed in the previous studies, the prolate case exhibits a different distribution from the previous works \citep[e.g., see figures 3 and 2 in][]{Matsumoto2018}. These results may provide a useful reference for interpreting observed continuum polarization and constraining the geometry of unresolved photospheres in future observations.

\subsection{General Triaxial Ellipsoids}
\begin{figure}
  \centering
  \includegraphics[width=\columnwidth, trim=2pt 2pt 2pt 2pt, clip]{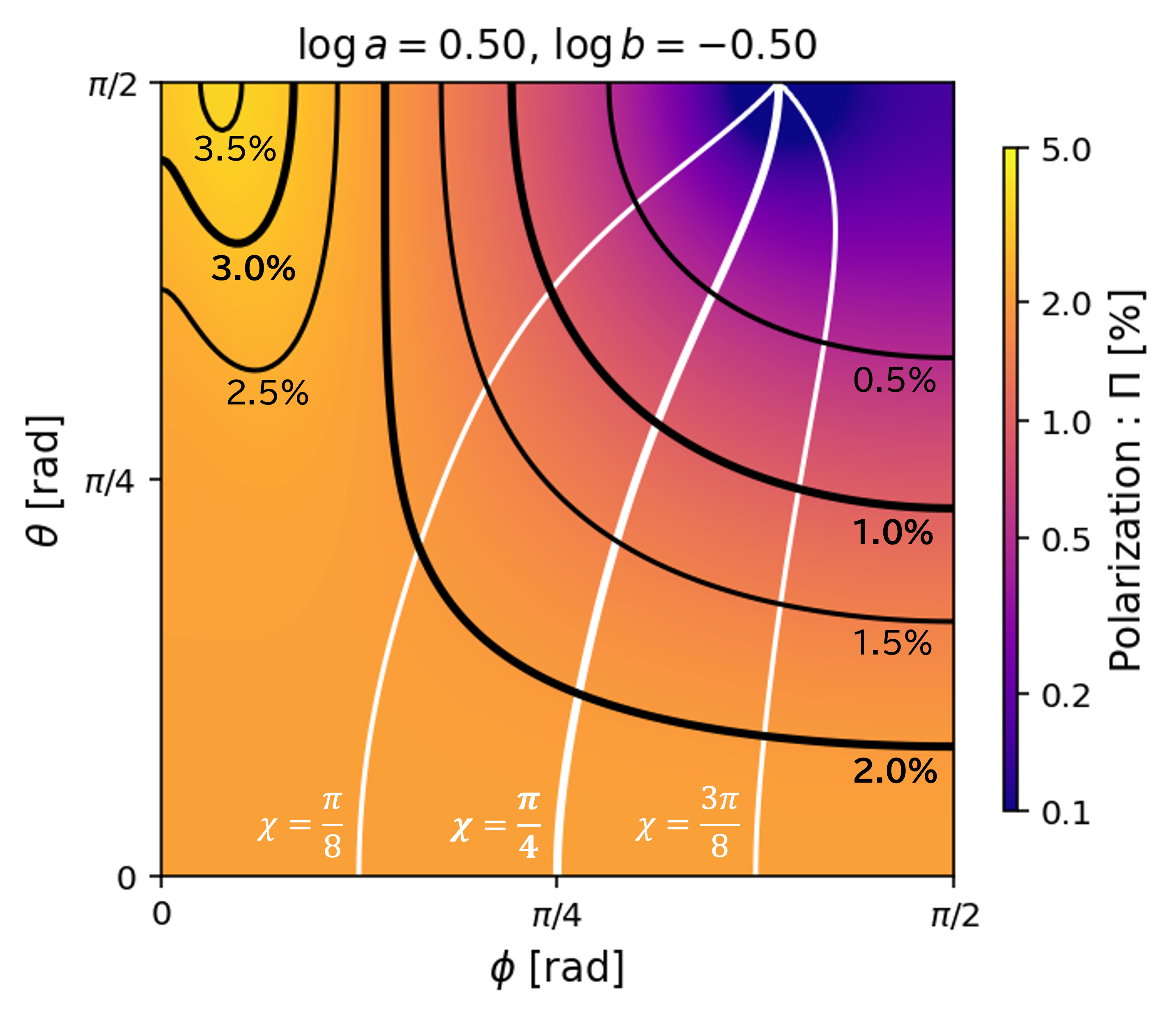}
  \caption{The degree of polarization in the $(\theta,\phi)$ plane for an ellipsoid with $\log a=0.5$ and $\log b=-0.5$. Black contours indicate constant polarization degrees at intervals of $0.5\,\%$, with thicker lines at integer percentages. White contours indicate constant polarization angles at intervals of $\pi/8$, with a thicker line at $\pi/4$. Polarization degrees outside the range $0.1\text{–}5\,\%$ are clipped to the corresponding limits of the color scale.}
  \label{fig:ellipsoid map1}
\end{figure}

We investigate the polarization signature from general triaxial ellipsoids based on the insight obtained for spheroids. For the general case, there are four parameters, ($\theta,\phi$) specify the viewing direction, and ($a,b$) specify the ellipsoidal shape. Given the relatively large numbers of free parameters, we begin by discussing the viewing-angle dependence of a triaxial ellipsoid elongated along the $x$-axis ($\log a=0.5$) and compressed along the $y$-axis ($\log b=-0.5$). Figure~\ref{fig:ellipsoid map1} shows a contour map of the polarization degree $\Pi$ and angle $\chi$ in the ($\phi,\theta$) plane.

\if{
\begin{figure}
  \centering
  \includegraphics[width=\columnwidth]{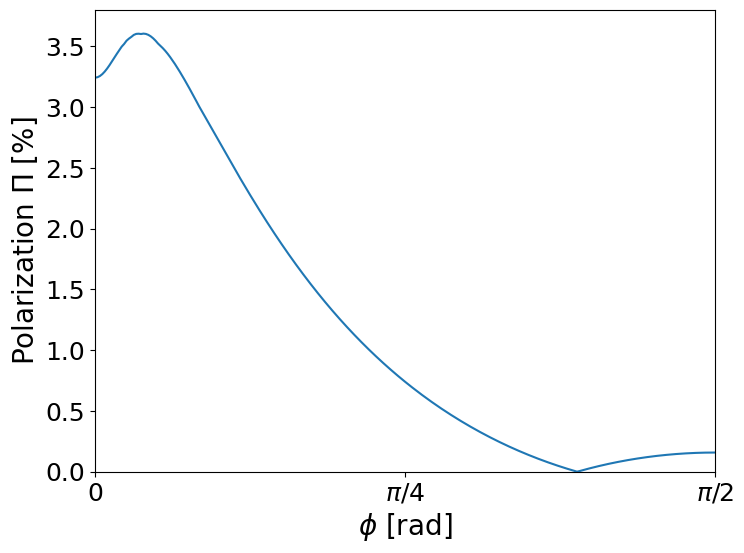}
  \caption{The polarization degree as a function of the azimuthal angle $\phi$ along the equatorial slice ($\theta=\pi/2$) of Fig.~\ref{fig:ellipsoid map1}. As in highly oblate spheroids, the polarization degree reaches its maximum at a viewing angle slightly offset from the x-axis rather than exactly along it. It also exhibits a null point between $\phi=\pi/4$ and $\pi/2$, where the polarization angle flips by $90^\circ$, analogous to the behavior of highly prolate spheroids.}
  \label{fig:ellipsoid pol}
\end{figure}
}\fi

For a given shape, the polarization degree attains its maximum and minimum when viewed along the $x$-axis ($\theta\simeq\pi/2$ and $\phi\simeq0$) and the $y$-axis ($\theta\simeq\pi/2$ and $\phi\simeq\pi/2$), respectively. This behavior can be understood by considering the area of surface elements viewed nearly edge-on ($\mu\sim0$). As demonstrated by the Chandrasekhar solution, polarization from a plane-parallel atmosphere is strongest when viewed edge-on \citep{Chandrasekhar1960}. This suggests that, as a simple geometric approximation, the degree of polarization roughly scales with the area of nearly edge-on surface elements.

We can therefore estimate the relative degree of polarization for different viewing directions by comparing these areas. For example, an oblate spheroid exhibits its maximum polarization near $\theta\simeq\pi/2$, where the large upper and lower surfaces of the spheroid are viewed nearly edge-on. For a general ellipsoid, we estimate these areas by approximating the ellipsoid as a rectangular box. When viewed along the $x$-, $y$-, and $z$-axes, the areas of the nearly edge-on surfaces scale as $S_x\sim a(b+c)$, $S_y\sim b(a+c)$, and $S_z\sim c(a+b)$, respectively. For the present axis ordering, $a>c=1>b$, these satisfy $S_x>S_z>S_y$. We thus expect the polarization to follow the same ordering. This agrees with the locations of the maximum and minimum polarization seen in Fig.~\ref{fig:ellipsoid map1} and, as shown below, remains a useful guide for more general configurations.

\begin{figure*}
\centering
\includegraphics[width=\textwidth]{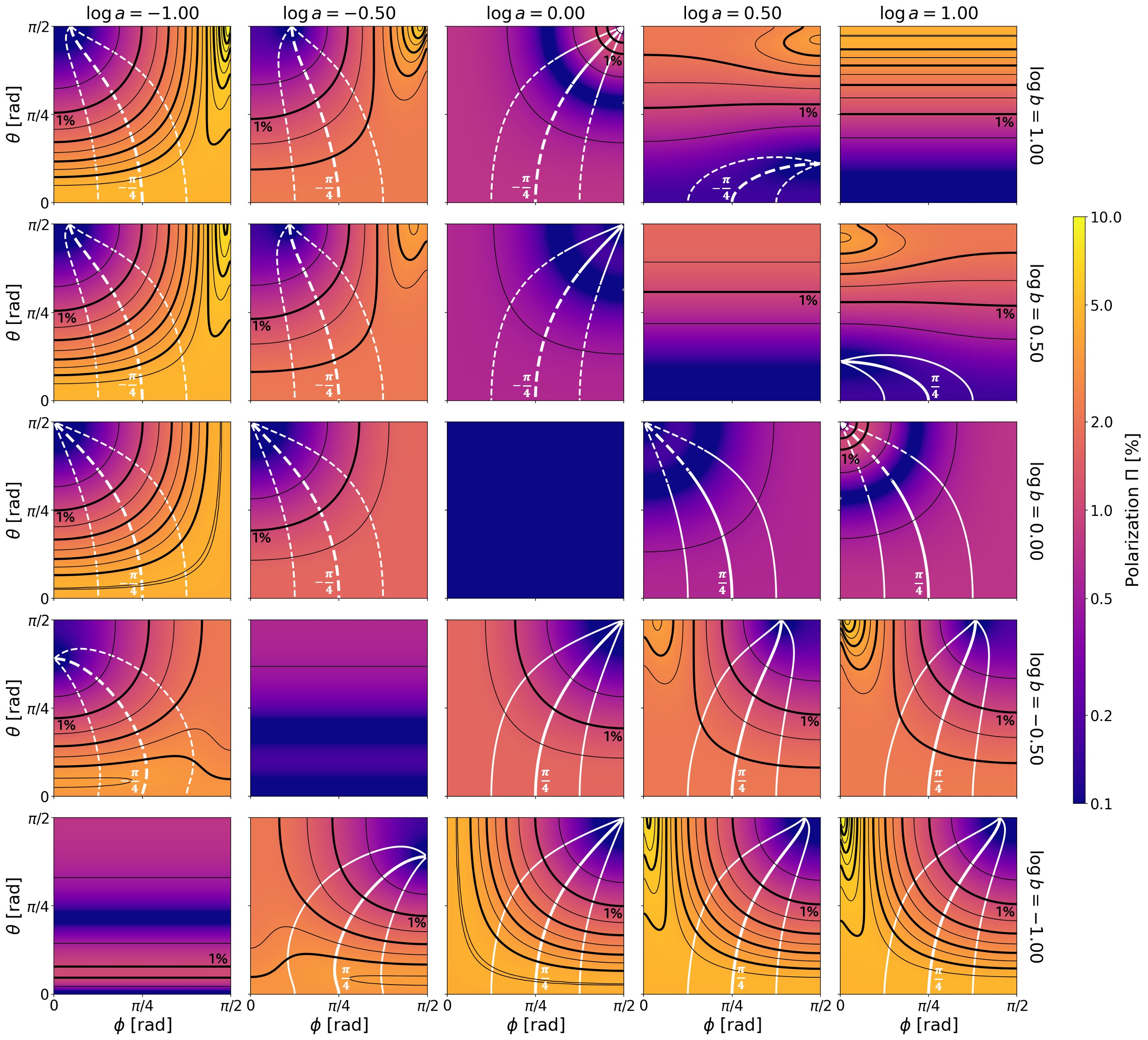}
\caption{Same as Fig.~\ref{fig:ellipsoid map1}, but for different ellipsoidal shapes specified by $(a,b)$, with $-1\leq\log a,\log b\leq1$. The white dashed contours indicate constant negative polarization angles.}
\label{fig:ellipsoid map}
\end{figure*}

Figure~\ref{fig:ellipsoid map} presents the polarization degree in the ($\phi,\theta$) plane for a wide range of ellipsoidal shapes. Each row and column correspond to a fixed value of $a$ and $b$, respectively, while the contours show the viewing-angle dependence of the polarization degree and angle as in Fig.~\ref{fig:ellipsoid map1}. The panels along the diagonal running from the lower-left to the upper-right corner correspond to spheroids ($a=b$). Panels located symmetrically with respect to this diagonal, such as those for $(\log a,\log b)=(0.5,-0.5)$ and $(-0.5,0.5)$, represent equivalent ellipsoidal shapes up to an interchange of the $a$- and $b$-axes, and therefore exhibit equivalent polarization distributions. In addition, the row of panels with $b=1$ and the column of panels with $a=1$ correspond to spheroids whose symmetry axes are the $x$- and $y$-axes, respectively.

For a given ellipsoidal shape, the polarization degree depends strongly on the viewing angle. As discussed above, the maximum and minimum polarization degrees are generally attained when the object is viewed along the principal axes for which the areas of nearly edge-on surface elements are largest and smallest, respectively. From an observational perspective, a key question is which photospheric geometries can reproduce a given observed polarization degree. For this purpose, it is useful to characterize the maximum polarization attainable for each ellipsoidal shape.

\begin{figure}
  \centering
  \includegraphics[width=\columnwidth, trim=2pt 2pt 2pt 2pt, clip]{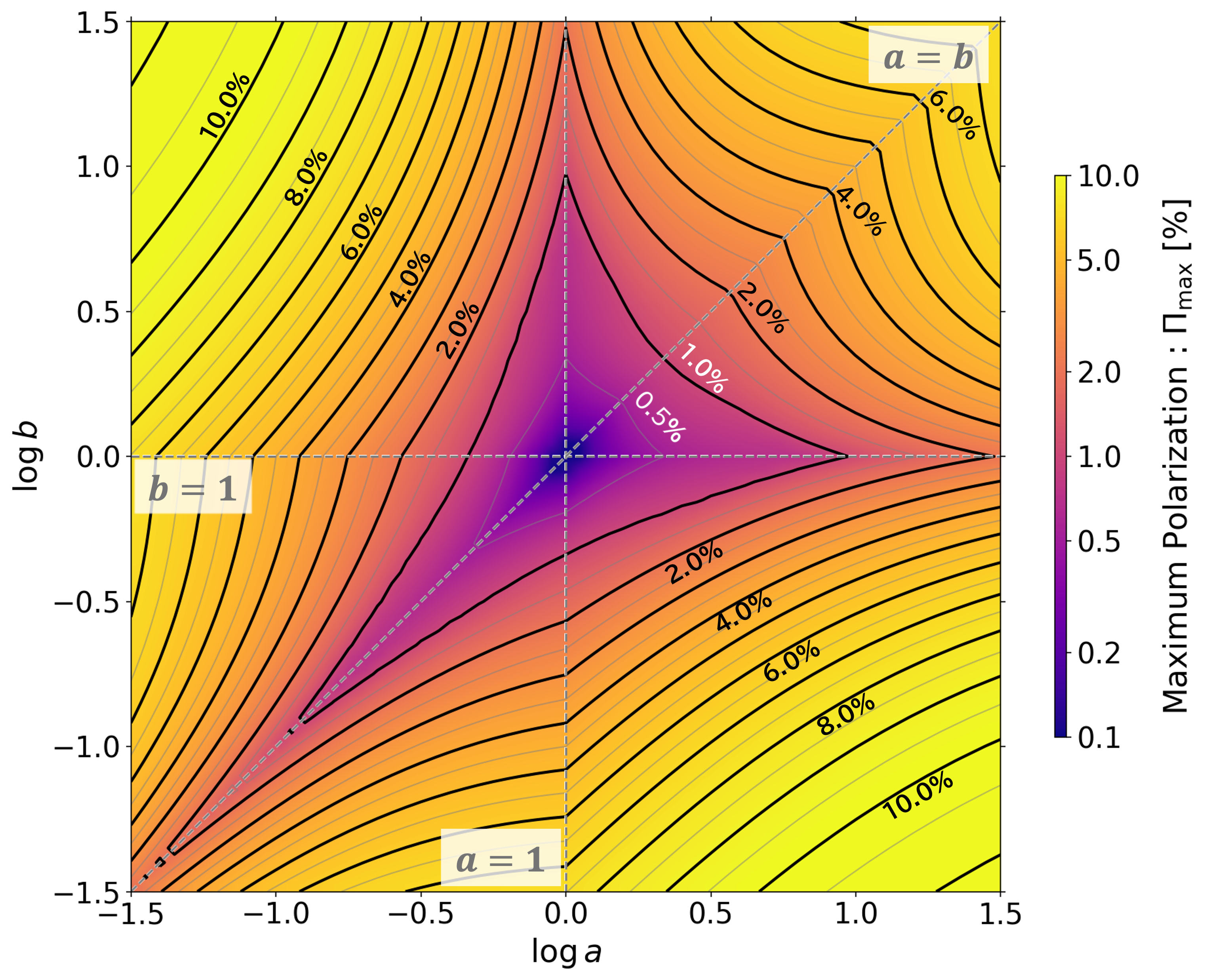}
  \caption{Maximum polarization over all viewing angles for each ellipsoidal shape. The thick contours indicate constant polarization degrees at intervals of $1\,\%$, while the thin contours are drawn at intermediate $0.5\,\%$ intervals. The dashed lines correspond to the three families of spheroids, namely, $a=b$, $a(=c)=1$, and $b(=c)=1$.
}
  \label{fig:polarization_max}
  \vspace{\baselineskip}
\end{figure}

Figure~\ref{fig:polarization_max} shows the maximum polarization degree attainable for each ellipsoidal geometry after varying the viewing direction over all ($\theta,\phi$). As in Fig.~\ref{fig:ellipsoid map}, the panels along the diagonal ($a=b$), as well as those along $a=1$ and $b=1$, correspond to spheroidal configurations. The distribution is symmetric with respect to the diagonal line ($a=b$), reflecting the fact that exchanging $a$ and $b$ leaves the ellipsoidal shape unchanged up to a relabeling of the coordinate axes.

Large polarization degrees, $\Pi\gtrsim5\,\%$, are achieved for highly unequal triaxial ellipsoids, corresponding to the upper-left ($a\ll c=1\ll b$) and lower-right ($b\ll c=1\ll a$) regions of the parameter space. In the extreme cases $(a,b)=(10,0.1)$ and $(0.1,10)$, the maximum polarization approaches the Chandrasekhar's limit for a plane-parallel atmosphere, $\Pi\simeq11.7\,\%$. This suggests that exploring even more extreme geometries is unlikely to significantly alter the achievable maximum polarization.

For a measured polarization degree, only those ellipsoidal geometries whose maximum polarization exceeds the observed value remain viable. Therefore, Fig.~\ref{fig:polarization_max} can be used to place direct constraints on the allowed axis ratios ($a,b$) from continuum-polarimetric observations.

\vspace{10pt}
\section{Summary} \label{sec:summary}

In this work, polarization signatures from Thomson-scattering-dominated photospheres have been explored systematically using the analytical framework of \cite{Shapiro&Sutherland1982}, over a broader range of photospheric geometries than previous studies.
We first revisited the polarization from axisymmetric spheroids by systematically exploring a wider range of deformations and viewing angles. We then generalized the calculation to fully triaxial ellipsoids. Although this generalization expands the parameter space to four dimensions, consisting of two axis ratios and two viewing angles, the resulting polarization properties can still be understood through simple geometric considerations. These computational results are being made publicly available to assist future research (see the DATA AVAILABILITY STATEMENT at the end of this paper). Our main findings are summarized as follows.

1. 
For spheroidal geometries with moderate deviations from spherical symmetry, the polarization increases with viewing angle approximately as $\Pi\propto\sin^2\theta$, consistent with previous studies. However, this scaling breaks down for more strongly deformed spheroids, for which the polarization exhibits a non-monotonic dependence on viewing angle. In particular, highly prolate spheroids show a characteristic double-peaked behavior, with the polarization vanishing at an intermediate angle where the polarization angle undergoes a $90^\circ$ flip. This behavior can be understood as a competition between polarization from the upper and elongated side surfaces (see Fig.~\ref{fig:picture}). At small but finite viewing angles, the upper surface dominates and produces polarization along the $y^\prime$-axis, as in the oblate case. As the viewing angle increases, the contribution from the elongated side surface, polarized along the $z^\prime$-axis, grows and eventually overtakes that from the upper surface. Their cancellation produces the polarization null, beyond which the polarization angle flips by $90^\circ$.

2.
For general triaxial ellipsoids, the polarization strongly depends on the viewing direction. The maximum polarization is generally attained when the ellipsoid is viewed along its longest principal axis. This configuration provides the largest area of surface elements viewed nearly edge-on ($\mu\simeq0$), where the local polarization is strongest \citep{Chandrasekhar1960}. The relative polarization for different viewing directions can therefore be understood, to first order, simply by comparing the areas of such nearly edge-on surface elements.

3.
By plotting the maximum polarization over the axis-ratio plane (Fig.~\ref{fig:polarization_max}), we find that, within the range $0.1<a,b<10$, the highest polarization degree of $\simeq10\,\%$ is achieved for highly unequal triaxial ellipsoids with $b\ll c=1\ll a$ or $a\ll c=1\ll b$. Such extreme geometries maximize the contrast between the surface areas viewed nearly edge-on along different viewing directions, thereby producing large net polarization. Interestingly, the resulting polarization degree is already close to the maximum polarization attainable for a plane-parallel atmosphere, $\simeq11.7\,\%$ \citep{Chandrasekhar1960}. This suggests that the polarization degree is nearly saturated for axis ratios of order $a/b\simeq100$, and that considering even more extreme geometries would not substantially increase the maximum polarization.

While our results can be understood in an intuitive manner, several caveats associated with our assumptions should be noted. One important assumption is that the radiative flux is uniform over the photosphere. This is likely a reasonable approximation for moderately aspherical geometries, but it becomes increasingly inaccurate as the deviation from spherical symmetry grows. Under the diffusion approximation, the radiative flux preferentially propagates toward directions where the gradient of the internal energy (or temperature) is steepest \citep[e.g.,][]{Rybicki&Lightman1979}, naturally leading to a non-uniform surface flux distribution. For instance, in highly prolate spheroids with a centrally located heating source, radiation tends to diffuse preferentially toward the equatorial regions, thereby enhancing the contribution of these regions to the net polarization. Constructing such a model requires specifying a physical model for the internal energy distribution, which is beyond the scope of this paper. A more realistic calculation incorporating physically motivated flux distributions is deferred to future work.

Another potential limitation is that our results ultimately rely on the plane-parallel atmosphere solution of \cite{Chandrasekhar1960}. We assume that the polarization produced by each surface element is determined locally. This assumption is justified when the optical-depth scale height is sufficiently small that both non-local radiative-transfer effects and the curvature of the photosphere can be neglected. However, the approximation may break down in more tenuous envelopes, where photons can travel over distances comparable to the size of the photosphere before escaping. In such cases, more sophisticated approaches, such as Monte Carlo radiative-transfer calculations, are required. Quantifying the regime in which the local plane-parallel approximation remains valid is left for future work.

Nonetheless, our calculations provide a physically intuitive understanding of polarization signatures from aspherical photospheres, which may serve as a useful guide for interpreting results from more sophisticated numerical simulations. In addition, the analytical framework developed here can readily be extended to accommodate more complicated and arbitrary geometries beyond idealized ellipsoids. We therefore expect that this work will provide a useful foundation for future studies aimed at extracting geometrical information from continuum-polarimetric observations of a wide variety of optical transients.

\vspace{1.0\baselineskip}
\section*{Acknowledgments}
N.H. thanks K. Koretomo for useful discussions. This
research was supported by JSPS KAKENHI 23H04900 and 24K17088 (TM), and
18K03692, 26H00822, 26H02063, and 26K22348 (TT).

\vspace{1.0\baselineskip}
\section*{Data availability statement}
The numerical data presented in this study are available on Zenodo at DOI: \href{https://doi.org/10.5281/zenodo.22538015}{10.5281/zenodo.22538015} \citep{Hachisu2026}.

\appendix
\section{Polarization from cylinder}
\label{appendix}
We calculate the polarization from a cylinder of height 2 (corresponding to $c=1$ of the ellipsoid) and radius $a$, which provides a useful approximation to the highly prolate ellipsoids discussed in \S\ref{sec:spheroids}. The contributions from the top and side surfaces are calculated separately.

For the top circular surface of the cylinder, the relevant quantities are
\begin{align}
&\bm{n}=(0,0,1)\ ,\\
&\mu=\cos\theta\ ,\\
&dA=\varpi d\varpi d\varphi\ ,\\
&\cos2\psi=1\ ,
\end{align}
where ($\varpi,\,\varphi$) denote the polar coordinates on the top circular surface. The net Stokes parameters from the top surface are then
\begin{align}
I_{\rm net}^{\rm top}&=\pi a^2 I(\cos\theta)\cos\theta\ ,\\
Q_{\rm net}^{\rm top}&=\pi a^2 Q(\cos\theta)\cos\theta \ .
\end{align}
We approximate the Chandrasekhar solution by the linear forms
\begin{align}
I(\mu)&=I_0+I_1\mu\ ,\\
Q(\mu)&=Q_0(\mu-1)\ ,
\end{align} where the fitting constants are $I_0=0.442$, $I_1=0.837$, and $Q_0=0.0424$.

For the side surface, the relevant quantities are
\begin{align}
&\bm{n}=(\cos\varphi,\sin\varphi,0)\ ,\\
&\mu=\sin\theta\cos\varphi\ ,\\
&dA=a\,dz\,d\varphi\ ,\\
&\cos2\psi= \frac{\cos^2\theta-\tan^2\varphi} {\cos^2\theta+\tan^2\varphi}\ .
\end{align}
The Stokes $I$ parameter is given by
\begin{align}
I_{\rm net}^{\rm side} = a \int_{-1}^1 dz \int_{-\pi/2}^{\pi/2} d\varphi\, \mu\left(I_0+I_1\mu\right)\ , \end{align}
where the azimuthal integration range is restricted to the visible side surface. Carrying out the integrals independently, we obtain
\begin{align}
I_{\rm net}^{\rm side} = 4a\sin\theta \left( I_0+\frac{\pi I_1}{4}\sin\theta \right)\ .
\end{align}
Similarly, the Stokes $Q$ parameter is given by
\begin{align}
Q_{\rm net}^{\rm side} &= aQ_0 \int_{-1}^1 dz \int_{-\pi/2}^{\pi/2} d\varphi\, \mu(\mu-1) \left( \frac{\cos^2\theta-\tan^2\varphi} {\cos^2\theta+\tan^2\varphi} \right)
    \nonumber\\
&= 4aQ_0 \Biggl[ \frac{\cos\theta}{\sin^2\theta} \left\{ \pi(1-\cos\theta)-2\theta \right\}
    \nonumber\\
&\hspace{3cm} +\frac{1+\cos^2\theta}{\sin\theta} \left( 1-\frac{\pi}{4}\sin\theta \right) \Biggr] \ . 
\end{align}
The total polarization is then obtained from
\begin{align}
I_{\rm net} &= I_{\rm net}^{\rm top} + I_{\rm net}^{\rm side}\ , \\
Q_{\rm net} &= Q_{\rm net}^{\rm top} + Q_{\rm net}^{\rm side}\ .
\end{align}

\bibliography{compiled_bibliography}{}

@ARTICLE{Tanaka+2009b,
       author = {{Tanaka}, Masaomi and {Kawabata}, Koji S. and {Maeda}, Keiichi and {Iye}, Masanori and {Hattori}, Takashi and {Pian}, Elena and {Nomoto}, Ken'ichi and {Mazzali}, Paolo A. and {Tominaga}, Nozomu},
        title = "{Spectropolarimetry of the Unique Type Ib Supernova 2005bf: Larger Asymmetry Revealed by Later-Phase Data}",
      journal = {\apj},
         year = 2009,
        month = jul,
       volume = {699},
       number = {2},
        pages = {1119-1124},
          doi = {10.1088/0004-637X/699/2/1119},
archivePrefix = {arXiv},
       eprint = {0906.1062},
 primaryClass = {astro-ph.HE},
       adsurl = {https://ui.adsabs.harvard.edu/abs/2009ApJ...699.1119T}
}

@ARTICLE{Tanaka+2008c,
       author = {{Tanaka}, Masaomi and {Kawabata}, Koji S. and {Maeda}, Keiichi and {Hattori}, Takashi and {Nomoto}, Ken'ichi},
        title = "{Optical Spectropolarimetry and Asphericity of the Type Ic SN 2007gr}",
      journal = {\apj},
         year = 2008,
        month = dec,
       volume = {689},
       number = {2},
        pages = {1191-1198},
          doi = {10.1086/592325},
archivePrefix = {arXiv},
       eprint = {0806.1589},
 primaryClass = {astro-ph},
       adsurl = {https://ui.adsabs.harvard.edu/abs/2008ApJ...689.1191T}
}

@ARTICLE{Maund+2007,
       author = {{Maund}, Justyn R. and {Wheeler}, J. Craig and {Patat}, Ferdinando and {Baade}, Dietrich and {Wang}, Lifan and {H{\"o}flich}, Peter},
        title = "{Spectropolarimetry of the Type Ib/c SN 2005bf}",
      journal = {\mnras},
         year = 2007,
        month = oct,
       volume = {381},
       number = {1},
        pages = {201-210},
          doi = {10.1111/j.1365-2966.2007.12230.x},
archivePrefix = {arXiv},
       eprint = {0707.2237},
 primaryClass = {astro-ph},
       adsurl = {https://ui.adsabs.harvard.edu/abs/2007MNRAS.381..201M}
}

@ARTICLE{Wang+1996b,
       author = {{Wang}, Lifan and {Wheeler}, J. Craig and {Li}, Zongwei and {Clocchiatti}, Alejandro},
        title = "{Broadband Polarimetry of Supernovae: SN 1994D, SN 1994Y, SN 1994ae, SN 1995D, and SN 1995H}",
      journal = {\apj},
         year = 1996,
        month = aug,
       volume = {467},
        pages = {435},
          doi = {10.1086/177617},
archivePrefix = {arXiv},
       eprint = {astro-ph/9602155},
 primaryClass = {astro-ph},
       adsurl = {https://ui.adsabs.harvard.edu/abs/1996ApJ...467..435W}
}

@ARTICLE{Leonard+2005,
       author = {{Leonard}, Douglas C. and {Li}, Weidong and {Filippenko}, Alexei V. and {Foley}, Ryan J. and {Chornock}, Ryan},
        title = "{Evidence for Spectropolarimetric Diversity in Type Ia Supernovae}",
      journal = {\apj},
         year = 2005,
        month = oct,
       volume = {632},
       number = {1},
        pages = {450-475},
          doi = {10.1086/432866},
archivePrefix = {arXiv},
       eprint = {astro-ph/0506470},
 primaryClass = {astro-ph},
       adsurl = {https://ui.adsabs.harvard.edu/abs/2005ApJ...632..450L}
}

@ARTICLE{Wang+2001,
       author = {{Wang}, Lifan and {Howell}, D. Andrew and {H{\"o}flich}, Peter and {Wheeler}, J. Craig},
        title = "{Bipolar Supernova Explosions}",
      journal = {\apj},
         year = 2001,
        month = apr,
       volume = {550},
       number = {2},
        pages = {1030-1035},
          doi = {10.1086/319822},
       adsurl = {https://ui.adsabs.harvard.edu/abs/2001ApJ...550.1030W}
}

@ARTICLE{Wang+2003b,
       author = {{Wang}, Lifan and {Baade}, Dietrich and {H{\"o}flich}, Peter and {Khokhlov}, Alexei and {Wheeler}, J. Craig and {Kasen}, D. and {Nugent}, Peter E. and {Perlmutter}, Saul and {Fransson}, Claes and {Lundqvist}, Peter},
        title = "{Spectropolarimetry of SN 2001el in NGC 1448: Asphericity of a Normal Type Ia Supernova}",
      journal = {\apj},
         year = 2003,
        month = jul,
       volume = {591},
       number = {2},
        pages = {1110-1128},
          doi = {10.1086/375444},
archivePrefix = {arXiv},
       eprint = {astro-ph/0303397},
 primaryClass = {astro-ph},
       adsurl = {https://ui.adsabs.harvard.edu/abs/2003ApJ...591.1110W}
}

@ARTICLE{Leonard&Filippenko2001,
       author = {{Leonard}, Douglas C. and {Filippenko}, Alexei V.},
        title = "{Spectropolarimetry of the Type II Supernovae 1997ds, 1998A, and 1999gi}",
      journal = {\pasp},
         year = 2001,
        month = aug,
       volume = {113},
       number = {786},
        pages = {920-936},
          doi = {10.1086/322151},
archivePrefix = {arXiv},
       eprint = {astro-ph/0105295},
 primaryClass = {astro-ph},
       adsurl = {https://ui.adsabs.harvard.edu/abs/2001PASP..113..920L}
}

@ARTICLE{Leonard+2001,
       author = {{Leonard}, Douglas C. and {Filippenko}, Alexei V. and {Ardila}, David R. and {Brotherton}, Michael S.},
        title = "{Is It Round? Spectropolarimetry of the Type II-p Supernova 1999EM}",
      journal = {\apj},
         year = 2001,
        month = jun,
       volume = {553},
       number = {2},
        pages = {861-885},
          doi = {10.1086/320959},
archivePrefix = {arXiv},
       eprint = {astro-ph/0009285},
 primaryClass = {astro-ph},
       adsurl = {https://ui.adsabs.harvard.edu/abs/2001ApJ...553..861L}
}

@ARTICLE{Kasen+2003,
       author = {{Kasen}, Daniel and {Nugent}, Peter and {Wang}, Lifan and {Howell}, D.~A. and {Wheeler}, J. Craig and {H{\"o}flich}, Peter and {Baade}, Dietrich and {Baron}, E. and {Hauschildt}, P.~H.},
        title = "{Analysis of the Flux and Polarization Spectra of the Type Ia Supernova SN 2001el: Exploring the Geometry of the High-Velocity Ejecta}",
      journal = {\apj},
         year = 2003,
        month = aug,
       volume = {593},
       number = {2},
        pages = {788-808},
          doi = {10.1086/376601},
archivePrefix = {arXiv},
       eprint = {astro-ph/0301312},
 primaryClass = {astro-ph},
       adsurl = {https://ui.adsabs.harvard.edu/abs/2003ApJ...593..788K}
}

@ARTICLE{JordanaMitjans+2025,
       author = {{Jordana-Mitjans}, N. and {Franckowiak}, A. and {Ram{\'\i}rez-Ruiz}, E. and {Mundell}, C.~G. and {Valtonen-Mattila}, N. and {Stein}, R. and {Mil{\'a}n Veres}, P.},
        title = "{Optical polarisation of stellar-fed active and quiescent supermassive black holes}",
      journal = {\aap},
         year = 2025,
        month = dec,
       volume = {704},
          eid = {A250},
        pages = {A250},
          doi = {10.1051/0004-6361/202556305},
archivePrefix = {arXiv},
       eprint = {2507.06296},
 primaryClass = {astro-ph.HE},
       adsurl = {https://ui.adsabs.harvard.edu/abs/2025A&A...704A.250J}
}

@ARTICLE{Liodakis+2023,
       author = {{Liodakis}, I. and {Koljonen}, K.~I.~I. and {Blinov}, D. and {Lindfors}, E. and {Alexander}, K.~D. and {Hovatta}, T. and {Berton}, M. and {Hajela}, A. and {Jormanainen}, J. and {Kouroumpatzakis}, K. and {Mandarakas}, N. and {Nilsson}, K.},
        title = "{Optical polarization from colliding stellar stream shocks in a tidal disruption event}",
      journal = {Science},
         year = 2023,
        month = may,
       volume = {380},
       number = {6645},
        pages = {656-658},
          doi = {10.1126/science.abj9570},
archivePrefix = {arXiv},
       eprint = {2208.14465},
 primaryClass = {astro-ph.HE},
       adsurl = {https://ui.adsabs.harvard.edu/abs/2023Sci...380..656L}
}

@ARTICLE{Leonard+2006,
       author = {{Leonard}, Douglas C. and {Filippenko}, Alexei V. and {Ganeshalingam}, Mohan and {Serduke}, Franklin J.~D. and {Li}, Weidong and {Swift}, Brandon J. and {Gal-Yam}, Avishay and {Foley}, Ryan J. and {Fox}, Derek B. and {Park}, Sung and {Hoffman}, Jennifer L. and {Wong}, Diane S.},
        title = "{A non-spherical core in the explosion of supernova SN 2004dj}",
      journal = {\nat},
         year = 2006,
        month = mar,
       volume = {440},
       number = {7083},
        pages = {505-507},
          doi = {10.1038/nature04558},
archivePrefix = {arXiv},
       eprint = {astro-ph/0603297},
 primaryClass = {astro-ph},
       adsurl = {https://ui.adsabs.harvard.edu/abs/2006Natur.440..505L}
}

@ARTICLE{Dessart&Hillier2011b,
       author = {{Dessart}, Luc and {Hillier}, D. John},
        title = "{Synthetic line and continuum linear-polarization signatures of axisymmetric Type II supernova ejecta}",
      journal = {\mnras},
         year = 2011,
        month = aug,
       volume = {415},
       number = {4},
        pages = {3497-3519},
          doi = {10.1111/j.1365-2966.2011.18967.x},
archivePrefix = {arXiv},
       eprint = {1104.5346},
 primaryClass = {astro-ph.SR},
       adsurl = {https://ui.adsabs.harvard.edu/abs/2011MNRAS.415.3497D}
}

@ARTICLE{Nagao+2026,
       author = {{Nagao}, T. and {Maeda}, K. and {Matsumoto}, T.},
        title = "{Analytical modeling of polarization signals arising from confined circumstellar material in Type II supernovae}",
      journal = {\aap},
         year = 2026,
        month = mar,
       volume = {707},
          eid = {A103},
        pages = {A103},
          doi = {10.1051/0004-6361/202557386},
archivePrefix = {arXiv},
       eprint = {2509.19874},
 primaryClass = {astro-ph.SR},
       adsurl = {https://ui.adsabs.harvard.edu/abs/2026A&A...707A.103N}
}

@INCOLLECTION{Patat2017,
       author = {{Patat}, Ferdinando},
        title = "{Introduction to Supernova Polarimetry}",
    publisher = {Springer},
    booktitle = {Handbook of Supernovae},
         year = 2017,
       editor = {{Alsabti}, Athem W. and {Murdin}, Paul},
        pages = {1017},
          doi = {10.1007/978-3-319-21846-5_110},
       adsurl = {https://ui.adsabs.harvard.edu/abs/2017hsn..book.1017P}
}

@ARTICLE{Pursiainen+2023,
       author = {{Pursiainen}, M. and {Leloudas}, G. and {Cikota}, A. and {Bulla}, M. and {Inserra}, C. and {Patat}, F. and {Wheeler}, J.~C. and {Aamer}, A. and {Gal-Yam}, A. and {Maund}, J. and {Nicholl}, M. and {Schulze}, S. and {Sollerman}, J. and {Yang}, Y.},
        title = "{Polarimetry of hydrogen-poor superluminous supernovae}",
      journal = {\aap},
         year = 2023,
        month = jun,
       volume = {674},
          eid = {A81},
        pages = {A81},
          doi = {10.1051/0004-6361/202345945},
archivePrefix = {arXiv},
       eprint = {2301.08111},
 primaryClass = {astro-ph.HE},
       adsurl = {https://ui.adsabs.harvard.edu/abs/2023A&A...674A..81P}
}

@ARTICLE{Nagao+2024,
       author = {{Nagao}, T. and {Patat}, F. and {Cikota}, A. and {Baade}, D. and {Mattila}, S. and {Kotak}, R. and {Kuncarayakti}, H. and {Bulla}, M. and {Ayala}, B.},
        title = "{Spectropolarimetry of Type II supernovae. II. Intrinsic supernova polarization and its relation to photometric and spectroscopic properties}",
      journal = {\aap},
         year = 2024,
        month = jan,
       volume = {681},
          eid = {A11},
        pages = {A11},
          doi = {10.1051/0004-6361/202346715},
archivePrefix = {arXiv},
       eprint = {2308.00996},
 primaryClass = {astro-ph.HE},
       adsurl = {https://ui.adsabs.harvard.edu/abs/2024A&A...681A..11N}
}

@ARTICLE{Koljonen+2025b,
       author = {{Koljonen}, Karri I.~I. and {Nilsson}, Kari and {Liodakis}, Ioannis and {Lindfors}, Elina},
        title = "{Optical polarization properties of the closest tidal disruption event AT 2023clx indicate origin from tidal stream shocks}",
      journal = {\mnras},
         year = 2025,
        month = sep,
       volume = {542},
       number = {3},
        pages = {2238-2244},
          doi = {10.1093/mnras/staf1340},
archivePrefix = {arXiv},
       eprint = {2508.09309},
 primaryClass = {astro-ph.HE},
       adsurl = {https://ui.adsabs.harvard.edu/abs/2025MNRAS.542.2238K}
}

@ARTICLE{Uno+2025,
       author = {{Uno}, Kohki and {Maeda}, Keiichi and {Nagao}, Takashi and {Leloudas}, Giorgos and {Charalampopoulos}, Panos and {Mattila}, Seppo and {Aoki}, Kentaro and {Taguchi}, Kenta and {Kawabata}, Miho and {Moldon}, Javier and {P{\'e}rez-Torres}, Miguel and {Pursiainen}, Miika and {Reynolds}, Thomas},
        title = "{Spectropolarimetry of a Nuclear Transient AT2023clx: Revealing the Geometrical Alignment between the Transient Outflow and the Nuclear Dusty Region}",
      journal = {\apjl},
         year = 2025,
        month = jun,
       volume = {986},
       number = {2},
          eid = {L23},
        pages = {L23},
          doi = {10.3847/2041-8213/add71b},
archivePrefix = {arXiv},
       eprint = {2503.19024},
 primaryClass = {astro-ph.HE},
       adsurl = {https://ui.adsabs.harvard.edu/abs/2025ApJ...986L..23U}
}

@ARTICLE{Singh+2024,
       author = {{Singh}, Avinash and {Teja}, Rishabh Singh and {Moriya}, Takashi J. and {Maeda}, Keiichi and {Kawabata}, Koji S. and {Tanaka}, Masaomi and {Imazawa}, Ryo and {Nakaoka}, Tatsuya and {Gangopadhyay}, Anjasha and {Yamanaka}, Masayuki and {Swain}, Vishwajeet and {Sahu}, D.~K. and {Anupama}, G.~C. and {Kumar}, Brajesh and {Anche}, Ramya M. and {Sano}, Yasuo and {Raj}, A. and {Agnihotri}, V.~K. and {Bhalerao}, Varun and {Bisht}, D. and {Bisht}, M.~S. and {Belwal}, K. and {Chakrabarti}, S.~K. and {Fujii}, Mitsugu and {Nagayama}, Takahiro and {Matsumoto}, Katsura and {Hamada}, Taisei and {Kawabata}, Miho and {Kumar}, Amit and {Kumar}, Ravi and {Malkan}, Brian K. and {Smith}, Paul and {Sakagami}, Yuta and {Taguchi}, Kenta and {Tominaga}, Nozomu and {Watanabe}, Arata},
        title = "{Unravelling the Asphericities in the Explosion and Multifaceted Circumstellar Matter of SN 2023ixf}",
      journal = {\apj},
         year = 2024,
        month = nov,
       volume = {975},
       number = {1},
          eid = {132},
        pages = {132},
          doi = {10.3847/1538-4357/ad7955},
archivePrefix = {arXiv},
       eprint = {2405.20989},
 primaryClass = {astro-ph.HE},
       adsurl = {https://ui.adsabs.harvard.edu/abs/2024ApJ...975..132S}
}

@ARTICLE{Dessart+2024b,
       author = {{Dessart}, Luc and {Hillier}, D. John and {Leonard}, Douglas C.},
        title = "{The evolution of continuum polarization in type II supernovae as a diagnostic of ejecta morphology}",
      journal = {\aap},
         year = 2024,
        month = apr,
       volume = {684},
          eid = {A16},
        pages = {A16},
          doi = {10.1051/0004-6361/202347808},
archivePrefix = {arXiv},
       eprint = {2401.07330},
 primaryClass = {astro-ph.SR},
       adsurl = {https://ui.adsabs.harvard.edu/abs/2024A&A...684A..16D}
}

@ARTICLE{Bilinski+2024,
       author = {{Bilinski}, Christopher and {Smith}, Nathan and {Williams}, G. Grant and {Smith}, Paul S. and {Leonard}, Douglas C. and {Hoffman}, Jennifer L. and {Andrews}, Jennifer E. and {Milne}, Peter},
        title = "{Multi-epoch spectropolarimetry for a sample of Type IIn Supernovae: persistent asymmetry in dusty circumstellar material}",
      journal = {\mnras},
         year = 2024,
        month = apr,
       volume = {529},
       number = {2},
        pages = {1104-1129},
          doi = {10.1093/mnras/stae380},
archivePrefix = {arXiv},
       eprint = {2304.13034},
 primaryClass = {astro-ph.HE},
       adsurl = {https://ui.adsabs.harvard.edu/abs/2024MNRAS.529.1104B}
}

@ARTICLE{Maund+2023,
       author = {{Maund}, Justyn R. and {H{\"o}flich}, Peter A. and {Steele}, Iain A. and {Yang(\UTF{6768}\UTF{8F76})}, Yi and {Wiersema}, Klaas and {Kobayashi}, Shiho and {Jordana-Mitjans}, Nuria and {Mundell}, Carole and {Gomboc}, Andreja and {Guidorzi}, Cristiano and {Smith}, Robert J.},
        title = "{A flash of polarized optical light points to an aspherical 'cow'}",
      journal = {\mnras},
         year = 2023,
        month = may,
       volume = {521},
       number = {3},
        pages = {3323-3332},
          doi = {10.1093/mnras/stad539},
archivePrefix = {arXiv},
       eprint = {2303.00787},
 primaryClass = {astro-ph.SR},
       adsurl = {https://ui.adsabs.harvard.edu/abs/2023MNRAS.521.3323M}
}

@ARTICLE{Shrestha+2025,
       author = {{Shrestha}, Manisha and {DeSoto}, Sabrina and {Sand}, David J. and {Williams}, G. Grant and {Hoffman}, Jennifer L. and {Smith}, Paul S. and {McCall}, Callum and {Maund}, Justyn R. and {Steele}, Iain A. and {Wiersema}, Klaas and {Andrews}, Jennifer E. and {Smith}, Nathan and {Bilinski}, Christopher and {Milne}, Peter and {Anche}, Ramya M. and {Bostroem}, K. Azalee and {Hosseinzadeh}, Griffin and {Pearson}, Jeniveve and {Leonard}, Douglas C. and {Hsu}, Brian and {Dong}, Yize and {Hoang}, Emily and {Janzen}, Daryl and {Jencson}, Jacob E. and {Jha}, Saurabh W. and {Lundquist}, M.~J. and {Mehta}, Darshana and {Retamal}, Nicol{\'a}s Meza and {Valenti}, Stefano and {Farah}, Joseph and {Howell}, D. Andrew and {McCully}, Curtis and {Newsome}, Megan and {Gonzalez}, Estefania Padilla and {Pellegrino}, Craig and {Terreran}, Giacomo},
        title = "{Spectropolarimetry of SN 2023ixf Reveals Both Circumstellar Material and an Aspherical Helium Core}",
      journal = {\apjl},
         year = 2025,
        month = mar,
       volume = {982},
       number = {1},
          eid = {L32},
        pages = {L32},
          doi = {10.3847/2041-8213/adbb63},
archivePrefix = {arXiv},
       eprint = {2410.08199},
 primaryClass = {astro-ph.HE},
       adsurl = {https://ui.adsabs.harvard.edu/abs/2025ApJ...982L..32S}
}

@ARTICLE{Mauerhan+2024,
       author = {{Mauerhan}, Jon C. and {Smith}, Nathan and {Williams}, G. Grant and {Smith}, Paul S. and {Filippenko}, Alexei V. and {Bilinski}, Christopher and {Zheng}, WeiKang and {Brink}, Thomas G. and {Hoffman}, Jennifer L. and {Leonard}, Douglas C. and {Milne}, Peter and {Jeffers}, Benjamin and {Modak}, Shaunak and {Stegman}, Samantha and {Zhang}, Keto D.},
        title = "{Record-breaking polarization from the interacting superluminous supernova 2017hcc}",
      journal = {\mnras},
         year = 2024,
        month = jan,
       volume = {527},
       number = {3},
        pages = {6090-6099},
          doi = {10.1093/mnras/stad3579},
archivePrefix = {arXiv},
       eprint = {2304.12368},
 primaryClass = {astro-ph.SR},
       adsurl = {https://ui.adsabs.harvard.edu/abs/2024MNRAS.527.6090M}
}

@ARTICLE{Vasylyev+2023,
       author = {{Vasylyev}, Sergiy S. and {Yang}, Yi and {Filippenko}, Alexei V. and {Patra}, Kishore C. and {Brink}, Thomas G. and {Wang}, Lifan and {Chornock}, Ryan and {Margutti}, Raffaella and {Gates}, Elinor L. and {Burgasser}, Adam J. and {Karpoor}, Preethi R. and {LeBaron}, Natalie and {Softich}, Emma and {Theissen}, Christopher A. and {Wiston}, Eli and {Zheng}, WeiKang},
        title = "{Early Time Spectropolarimetry of the Aspherical Type II Supernova SN 2023ixf}",
      journal = {\apjl},
         year = 2023,
        month = oct,
       volume = {955},
       number = {2},
          eid = {L37},
        pages = {L37},
          doi = {10.3847/2041-8213/acf1a3},
archivePrefix = {arXiv},
       eprint = {2307.01268},
 primaryClass = {astro-ph.HE},
       adsurl = {https://ui.adsabs.harvard.edu/abs/2023ApJ...955L..37V}
}

@ARTICLE{Vasylyev+2026,
       author = {{Vasylyev}, Sergiy S. and {Dessart}, Luc and {Yang}, Yi and {Filippenko}, Alexei V. and {Patra}, Kishore C. and {Brink}, Thomas G. and {Wang}, Lifan and {Chornock}, Ryan and {Margutti}, Raffaella and {Gates}, Elinor L. and {Burgasser}, Adam J. and {Sears}, Huei and {Karpoor}, Preethi R. and {LeBaron}, Natalie and {Softich}, Emma and {Theissen}, Christopher A. and {Wiston}, Eli and {Zheng}, WeiKang},
        title = "{Spectropolarimetric Evolution of SN 2023ixf: An Asymmetric Explosion in a Confined Aspherical Circumstellar Medium}",
      journal = {\apj},
         year = 2026,
        month = mar,
       volume = {1000},
       number = {1},
          eid = {18},
        pages = {18},
          doi = {10.3847/1538-4357/ae323f},
archivePrefix = {arXiv},
       eprint = {2505.03975},
 primaryClass = {astro-ph.HE},
       adsurl = {https://ui.adsabs.harvard.edu/abs/2026ApJ..1000...18V}
}

@ARTICLE{Matsumoto2018,
       author = {{Matsumoto}, Tatsuya},
        title = "{Polarization of the first-hour macronovae}",
      journal = {\mnras},
         year = 2018,
        month = nov,
       volume = {481},
       number = {1},
        pages = {1008-1015},
          doi = {10.1093/mnras/sty2317},
archivePrefix = {arXiv},
       eprint = {1807.04766},
 primaryClass = {astro-ph.HE},
       adsurl = {https://ui.adsabs.harvard.edu/abs/2018MNRAS.481.1008M}
}

@ARTICLE{Uno+2023b,
       author = {{Uno}, Kohki and {Nagao}, Takashi and {Maeda}, Keiichi and {Kuncarayakti}, Hanindyo and {Tanaka}, Masaomi and {Kawabata}, Koji S. and {Nakaoka}, Tatsuya and {Kawabata}, Miho and {Yamanaka}, Masayuki and {Aoki}, Kentaro and {Isogai}, Keisuke and {Ogawa}, Mao and {Tajitsu}, Akito and {Imazawa}, Ryo},
        title = "{SN 2020uem: a Possible Thermonuclear Explosion within a Dense Circumstellar Medium (II). The Properties of the CSM from Polarimetry and Light-curve Modeling}",
      journal = {\apj},
         year = 2023,
        month = feb,
       volume = {944},
       number = {2},
          eid = {204},
        pages = {204},
          doi = {10.3847/1538-4357/acb5eb},
archivePrefix = {arXiv},
       eprint = {2301.09922},
 primaryClass = {astro-ph.HE},
       adsurl = {https://ui.adsabs.harvard.edu/abs/2023ApJ...944..204U}
}

@ARTICLE{Wichern+2026,
       author = {{Wichern}, H.~C.~I. and {Leloudas}, G. and {Pursiainen}, M. and {Cikota}, A. and {Jaisawal}, G.~K. and {Charalampopoulos}, P. and {Bulla}, M. and {Dai}, L. and {Anderson}, J.~P. and {Gromadzki}, M. and {Guti{\'e}rrez}, C.~P. and {M{\"u}ller-Bravo}, T.~E. and {Nicholl}, M.},
        title = "{Revisiting the unification of tidal disruption events with polarimetry}",
      journal = {\aap},
         year = 2026,
        month = jan,
       volume = {705},
          eid = {A250},
        pages = {A250},
          doi = {10.1051/0004-6361/202556270},
archivePrefix = {arXiv},
       eprint = {2510.27007},
 primaryClass = {astro-ph.HE},
       adsurl = {https://ui.adsabs.harvard.edu/abs/2026A&A...705A.250W}
}

@ARTICLE{Patra+2022,
       author = {{Patra}, Kishore C. and {Lu}, Wenbin and {Brink}, Thomas G. and {Yang}, Yi and {Filippenko}, Alexei V. and {Vasylyev}, Sergiy S.},
        title = "{Spectropolarimetry of the tidal disruption event AT 2019qiz: a quasi-spherical reprocessing layer}",
      journal = {\mnras},
         year = 2022,
        month = sep,
       volume = {515},
       number = {1},
        pages = {138-145},
          doi = {10.1093/mnras/stac1727},
archivePrefix = {arXiv},
       eprint = {2206.09039},
 primaryClass = {astro-ph.HE},
       adsurl = {https://ui.adsabs.harvard.edu/abs/2022MNRAS.515..138P}
}

@ARTICLE{Cikota+2019,
       author = {{Cikota}, Aleksandar and {Patat}, Ferdinando and {Wang}, Lifan and {Wheeler}, J. Craig and {Bulla}, Mattia and {Baade}, Dietrich and {H{\"o}flich}, Peter and {Cikota}, Stefan and {Clocchiatti}, Alejandro and {Maund}, Justyn R. and {Stevance}, Heloise F. and {Yang}, Yi},
        title = "{Linear spectropolarimetry of 35 Type Ia supernovae with VLT/FORS: an analysis of the Si II line polarization}",
      journal = {\mnras},
         year = 2019,
        month = nov,
       volume = {490},
       number = {1},
        pages = {578-599},
          doi = {10.1093/mnras/stz2322},
archivePrefix = {arXiv},
       eprint = {1908.07526},
 primaryClass = {astro-ph.HE},
       adsurl = {https://ui.adsabs.harvard.edu/abs/2019MNRAS.490..578C}
}

@ARTICLE{Leonard+2000,
       author = {{Leonard}, Douglas C. and {Filippenko}, Alexei V. and {Barth}, Aaron J. and {Matheson}, Thomas},
        title = "{Evidence for Asphericity in the Type IIN Supernova SN 1998S}",
      journal = {\apj},
         year = 2000,
        month = jun,
       volume = {536},
       number = {1},
        pages = {239-254},
          doi = {10.1086/308910},
archivePrefix = {arXiv},
       eprint = {astro-ph/9908040},
 primaryClass = {astro-ph},
       adsurl = {https://ui.adsabs.harvard.edu/abs/2000ApJ...536..239L}
}

@ARTICLE{Leloudas+2022,
       author = {{Leloudas}, Giorgos and {Bulla}, Mattia and {Cikota}, Aleksandar and {Dai}, Lixin and {Thomsen}, Lars L. and {Maund}, Justyn R. and {Charalampopoulos}, Panos and {Roth}, Nathaniel and {Arcavi}, Iair and {Auchettl}, Katie and {Malesani}, Daniele B. and {Nicholl}, Matt and {Ramirez-Ruiz}, Enrico},
        title = "{An asymmetric electron-scattering photosphere around optical tidal disruption events}",
      journal = {Nature Astronomy},
         year = 2022,
        month = sep,
       volume = {6},
        pages = {1193-1202},
          doi = {10.1038/s41550-022-01767-z},
archivePrefix = {arXiv},
       eprint = {2207.06855},
 primaryClass = {astro-ph.HE},
       adsurl = {https://ui.adsabs.harvard.edu/abs/2022NatAs...6.1193L}
}

@ARTICLE{Koljonen+2024c,
       author = {{Koljonen}, Karri I.~I. and {Liodakis}, Ioannis and {Lindfors}, Elina and {Nilsson}, Kari and {Reynolds}, Thomas M. and {Charalampopoulos}, Panos and {Kouroumpatzakis}, Konstantinos and {McCall}, Callum and {Jermak}, Helen E. and {Steele}, Iain A. and {Carbajo-Hijarrubia}, Juan},
        title = "{The extreme coronal line emitter AT 2022fpx: varying optical polarization properties and late-time X-ray flare}",
      journal = {\mnras},
         year = 2024,
        month = jul,
       volume = {532},
       number = {1},
        pages = {112-125},
          doi = {10.1093/mnras/stae1466},
archivePrefix = {arXiv},
       eprint = {2403.04877},
 primaryClass = {astro-ph.HE},
       adsurl = {https://ui.adsabs.harvard.edu/abs/2024MNRAS.532..112K}
}

@ARTICLE{Floris+2025b,
       author = {{Floris}, A. and {Liodakis}, I. and {Koljonen}, K.~I.~I. and {Lindfors}, E. and {Ag{\'\i}s-Gonzalez}, B. and {Paggi}, A. and {Blinov}, D. and {Nilsson}, K. and {Agudo}, I. and {Charalampopoulos}, P. and {D{\'\i}az Teodori}, M.~A. and {Escudero Pedrosa}, J. and {Otero-Santos}, J. and {Piirola}, V. and {Newsome}, M. and {Van Velzen}, S.},
        title = "{Polarimetric diversity in tidal disruption events: Comparative study of low-polarised sources with AT2020mot}",
      journal = {\aap},
         year = 2025,
        month = nov,
       volume = {703},
          eid = {A81},
        pages = {A81},
          doi = {10.1051/0004-6361/202555626},
archivePrefix = {arXiv},
       eprint = {2509.08895},
 primaryClass = {astro-ph.HE},
       adsurl = {https://ui.adsabs.harvard.edu/abs/2025A&A...703A..81F}
}

@ARTICLE{Inserra+2016b,
       author = {{Inserra}, C. and {Bulla}, M. and {Sim}, S.~A. and {Smartt}, S.~J.},
        title = "{Spectropolarimetry of Superluminous Supernovae: Insight into Their Geometry}",
      journal = {\apj},
         year = 2016,
        month = nov,
       volume = {831},
       number = {1},
          eid = {79},
        pages = {79},
          doi = {10.3847/0004-637X/831/1/79},
archivePrefix = {arXiv},
       eprint = {1607.02353},
 primaryClass = {astro-ph.HE},
       adsurl = {https://ui.adsabs.harvard.edu/abs/2016ApJ...831...79I}
}

@ARTICLE{Lee+2020,
       author = {{Lee}, Chien-Hsiu and {Hung}, Tiara and {Matheson}, Thomas and {Soraisam}, Monika and {Narayan}, Gautham and {Saha}, Abhjita and {Stubens}, Carl and {Wolf}, Nicholas},
        title = "{Optical Polarimetry of the Tidal Disruption Event AT2019DSG}",
      journal = {\apjl},
         year = 2020,
        month = mar,
       volume = {892},
       number = {1},
          eid = {L1},
        pages = {L1},
          doi = {10.3847/2041-8213/ab7cd3},
       adsurl = {https://ui.adsabs.harvard.edu/abs/2020ApJ...892L...1L}
}

@ARTICLE{Chornock+2010b,
       author = {{Chornock}, Ryan and {Filippenko}, Alexei V. and {Li}, Weidong and {Silverman}, Jeffrey M.},
        title = "{Large Late-Time Asphericities in Three Type IIP Supernovae}",
      journal = {\apj},
         year = 2010,
        month = apr,
       volume = {713},
       number = {2},
        pages = {1363-1375},
          doi = {10.1088/0004-637X/713/2/1363},
archivePrefix = {arXiv},
       eprint = {0912.2465},
 primaryClass = {astro-ph.CO},
       adsurl = {https://ui.adsabs.harvard.edu/abs/2010ApJ...713.1363C}
}

@article{Brown&McLean1977,
	adsurl = {http://adsabs.harvard.edu/abs/1977A%26A....57..141B},
	author = {{Brown}, J.~C. and {McLean}, I.~S.},
	journal = {\aap},
	month = may,
	pages = {141},
	title = {{Polarisation by Thomson Scattering in Optically Thin Stellar Envelopes. I. Source Star at Centre of Axisymmetric Envelope}},
	volume = 57,
	year = 1977}

@ARTICLE{Kumar+2016,
       author = {{Kumar}, Brajesh and {Pandey}, S.~B. and {Eswaraiah}, C. and {Kawabata}, K.~S.},
        title = "{Broad-band polarimetric investigation of the Type II-plateau supernova 2013ej}",
      journal = {\mnras},
         year = 2016,
        month = mar,
       volume = {456},
       number = {3},
        pages = {3157-3167},
          doi = {10.1093/mnras/stv2720},
archivePrefix = {arXiv},
       eprint = {1511.05736},
 primaryClass = {astro-ph.HE},
       adsurl = {https://ui.adsabs.harvard.edu/abs/2016MNRAS.456.3157K}
}

@article{Wang&Wheeler2008,
	adsurl = {http://adsabs.harvard.edu/abs/2008ARA%26A..46..433W},
	archiveprefix = {arXiv},
	author = {{Wang}, L. and {Wheeler}, J.~C.},
	doi = {10.1146/annurev.astro.46.060407.145139},
	eprint = {0811.1054},
	journal = {\araa},
	month = sep,
	pages = {433-474},
	title = {{Spectropolarimetry of Supernovae}},
	volume = 46,
	year = 2008}

@article{Hoflich1991,
	adsurl = {http://ads.nao.ac.jp/abs/1991A%26A...246..481H},
	author = {{Hoflich}, P.},
	journal = {\aap},
	month = jun,
	pages = {481},
	title = {{Asphericity Effects in Scattering Dominated Photospheres}},
	volume = 246,
	year = 1991}

@book{Chandrasekhar1960,
  author    = {{Chandrasekhar}, S.},
  title     = {{Radiative Transfer}},
  publisher = {Dover},
  address   = {New York},
  year      = {1960},
}

@article{Shapiro&Sutherland1982,
	adsurl = {http://ads.nao.ac.jp/abs/1982ApJ...263..902S},
	author = {{Shapiro}, P.~R. and {Sutherland}, P.~G.},
	doi = {10.1086/160559},
	journal = {\apj},
	month = dec,
	pages = {902-924},
	title = {{The polarization of supernova light - A measure of deviation from spherical symmetry}},
	volume = 263,
	year = 1982}

@book{Rybicki&Lightman1979,
  author    = {{Rybicki}, G.~B. and {Lightman}, A.~P.},
  title     = {{Radiative Processes in Astrophysics}},
  publisher = {Wiley},
  address   = {New York},
  year      = {1979},
}

@misc{Hachisu2026,
  author = {{Hachisu}, Naoki. and {Matsumoto}, Tatsuya. and {Totani}, Tomonori.},
  title  = {Numerical results of polarization calculations for "Polarization Properties of Thomson-Scattering Triaxial Ellipsoids"},
  year   = {2026},
  publisher    = {Zenodo},
  version      = {v1.0},
  doi          = {10.5281/zenodo.22538015},
  url          = {https://doi.org/10.5281/zenodo.22538015}
}
\bibliographystyle{mnras}

\end{document}